\documentclass{aa}  

\usepackage{graphicx}
\usepackage{txfonts}
\usepackage{siunitx}
\usepackage{hyperref}
\usepackage{comment}
\usepackage{siunitx}
\newcommand{\matchcal}[1]{\mathcal{#1}}
\usepackage{float}
\usepackage{dblfloatfix}
\usepackage{placeins}
                                
\begin{document}

   \title{Galaxy clusters in the LoTSS-DR3: \\Catalogues and detection pipeline for diffuse radio emission}
   \titlerunning{Galaxy clusters in the LoTSS-DR3}
   \authorrunning{C. Stuardi et al.}


   \author{C. Stuardi\inst{1}, G. Di Gennaro\inst{1}, A. Botteon\inst{1}, F. Braga\inst{2}, C. Gheller\inst{1}, F. Vazza\inst{2}, M. Balboni\inst{2,3}, N. Biava\inst{1,6}, A.~Bonafede\inst{2,1}, M. Br\"uggen\inst{4}, G. Brunetti\inst{1}, R. Cassano\inst{1}, M. Cianfaglione\inst{2,1}, V. Cuciti\inst{2,1}, F. De Gasperin\inst{1}, F.~Gastaldello\inst{3}, M.J.~Hardcastle\inst{5}, M. Hoeft\inst{6}, H.J.A. Rottgering\inst{7}, N. Sanvitale\inst{1}, T. W. Shimwell\inst{8,7} \and R.J.~van~Weeren\inst{7}}

   \institute{INAF - Istituto di Radioastronomia (IRA), Via Gobetti 101, 40129 Bologna, Italy\\
             \email{ccstuardi@gmail.com}
             \and Dipartimento di Fisica e Astronomia, Universit\'{a} di Bologna, Via P. Gobetti 92/3, 40129 Bologna, Italy
            \and INAF - IASF Milano, via A. Corti 12, 20133 Milano, Italy
            \and Hamburger Sternwarte, University of Hamburg, Gojenbergsweg 112, 21029 Hamburg, Germany
            \and Department of Physics, Astronomy and Mathematics, University of Hertfordshire, College Lane, Hatfield AL10 9AB, UK
            \and Th\"uringer Landessternwarte, Sternwarte 5, 07778 Tautenburg, Germany
            \and Leiden Observatory, Leiden University, PO Box 9513, 2300 RA Leiden, The Netherlands
            \and ASTRON, Netherlands Institute for Radio Astronomy, Oude Hoogeveensedijk 4, 7991 PD, Dwingeloo, The Netherlands \\ }

   \date{Received XX}
 
  \abstract
{The third data release (DR3) of the LOFAR Two-metre Sky Survey (LoTSS) provides an unprecedented view of the northern sky at 144 MHz, containing more than 13 million radio sources. While compact sources can be efficiently identified with automated software packages, the detection of diffuse radio emission associated with galaxy clusters still requires dedicated processing and visual inspection. Given the scale of current and forthcoming radio surveys, automated approaches based on artificial intelligence are becoming essential to the identification of the most interesting targets.}  
{We aim to develop an automated pipeline to construct a catalogue of galaxy clusters hosting diffuse radio emission from LoTSS-DR3 20$\arcsec$ images. The pipeline is designed to provide both the probability that a cluster hosts diffuse radio emission and an interpretable image of its shape and morphology.}  
{We employed Radio U-Net, a convolutional neural network optimised for image segmentation (i.e. pixel-level identification) of diffuse radio emission. To associate detected emission with individual clusters, we combined the network output with positional, mass, and redshift information from four X-ray- and Sunyaev–Zeldovich-selected cluster catalogues, resulting in a merged sample of 3822 clusters covered by the LoTSS-DR3.}  
{We produced a pixel-level segmentation map of the full LoTSS-DR3 and a quantitative indicator ($\matchcal{R}$ value) for the presence of diffuse radio emission in each cluster. This enables the selection of sub-samples with specific properties for targeted follow-up or statistical studies. As a demonstration of the first application, we identified a sub-sample of 357 clusters selected at the highest network accuracy ($76\%$), and we showed some examples of newly detected systems. For the second, using a larger statistical sample, we verified that the detection fraction of diffuse radio sources in the four catalogues increases with the mass and redshift of
the clusters.}
{This work establishes a flexible and scalable framework for deep learning-based searches for diffuse radio emission in new-generation wide-area radio surveys.}

   \keywords{galaxies: clusters: intracluster medium -- techniques: image processing -- software: data analysis}

   \maketitle
   \nolinenumbers

\section{Introduction}

Galaxy clusters represent the largest gravitationally bound structures in the Universe and form at the nodes of the cosmic web, a vast network of filaments on megaparsec scales through which matter accretes. The thermal plasma that fills galaxy clusters, known as the intracluster medium (ICM), is permeated by weak magnetic fields \citep[typically 0.1–10 $\mu$G, see][for a review]{Govoni04}. During structure formation, mergers and accretion processes drive shocks and turbulence that can accelerate particles to relativistic energies \citep{Bruggen12,Brunetti14}, producing synchrotron radiation detectable at radio wavelengths and thereby revealing the presence of magnetic fields and relativistic electrons in the ICM.

Diffuse radio sources in galaxy clusters -- such as giant radio halos, mini-halos, radio relics, and revived fossil plasma sources -- are key tracers of these non-thermal components \citep{vanWeeren19}. Historically, these sources have been identified in surveys and then studied through targeted observations of individual systems \citep[e.g.][]{Feretti12}. In recent years, the advent of new-generation low-frequency interferometers such as the Low Frequency Array \citep[LOFAR;][]{vanHaarlem13}, MeerKAT \citep{Jonas16}, the Australian Kilometre Array Pathfinder \citep[ASKAP;][]{Hotan21}, and the Murchison Widefield Array \citep[MWA;][]{Tingay13} has dramatically improved our capability to detect faint and extended radio emission. Diffuse radio sources typically have a surface brightness of $S_{\rm 1.4~ GHz} < 1$ $\mu$Jy arcsec$^{-2}$ at 1.4 GHz and become brighter at lower observing frequencies ($\nu$) due to their steep radio spectrum ($\alpha > 1$, with $S_{\nu} \propto \nu^{-\alpha}$).

In particular, LOFAR was used to perform a systematic search for diffuse cluster radio emission at frequencies below 200 MHz in a mass-selected sample of 309 clusters \citep{Botteon22}. This work has provided the largest collection to date, listing 138 clusters with diffuse radio emission (radio relics, radio halos, or sources with uncertain classification) and allowed subsequent statistical studies aimed at understanding the origin of the detected sources \citep{Cuciti23, Cassano23, Jones23, Bruno23a}. Other recent collections include the MeerKAT Galaxy Cluster Legacy Survey \citep{Knowles22,Kolokythas25}, with 62 galaxy clusters hosting diffuse radio emission, and a pilot collection of 37 clusters in the Evolutionary Map of the Universe (EMU), performed with ASKAP \citep{Duchesne24}. With the upcoming Square Kilometre Array \citep[SKA;][]{Dewdney09}, the number of known diffuse sources is expected to increase by orders of magnitude \citep[e.g.][]{Nuza12,Cassano15}. More quantitatively, SKA1-Low is predicted to detect up to $\sim$2600 radio halos out to $z\sim0.6$ (Cassano et al. AASKAII), while the approximate number of detected radio relics will be $\sim$1000 (Pal et al. AASKAII) in the southern hemisphere. This increase will require robust automated methodologies for the detection of these sources.

Traditional approaches to identifying diffuse radio emission in galaxy clusters have typically relied on manual or highly customised data reduction procedures and visual inspection \citep[see e.g.][]{Giovannini20,Botteon22,Knowles22}. However, given the scale of modern surveys, these methods are no longer feasible, owing both to the increase in the number of sources and to the computational cost and effort required for extensive data reprocessing. Robust identification often requires cross-matching with multi-wavelength data (optical, X-ray, and Sunyaev–Zeldovich observations) to confirm associations with galaxy clusters, while spectro-polarimetric analyses are necessary in many cases to firmly establish the nature of the detected sources. While human validation will likely remain essential, the development of automated, robust, and efficient detection techniques is crucial to fully exploit the scientific potential of upcoming high-volume datasets.

In recent years, deep learning (DL) techniques have emerged as powerful tools for astronomical image analysis, enabling automated source detection, classification, and segmentation (here, segmentation refers to the pixel-by-pixel identification of regions associated with specific astrophysical emission). Convolutional neural networks have been successfully applied to radio astronomical images, in particular to solve the task of source detection \citep[e.g.][]{Lukic19, Riggi23, Sortino23, Cornu24} and for the classification of radio galaxies \citep[e.g.][]{Riggi24,Gupta24a,Lao25}. LOFAR images of radio galaxies have also been analysed using self-supervised learning \citep{Mostert21, Mostert23, BaronPerez25}. 

The application of DL techniques to the detection and segmentation of diffuse radio emission is still relatively unexplored, primarily due to the limited number of known sources. Supervised networks require large training datasets, which, in the absence of sufficiently extensive observational samples, can be supplemented with simulated data. Early applications of DL to the recognition of diffuse radio emission in cosmological simulations were presented by \citet{Gheller18}, while \citet{Gheller22} developed DL methods for the denoising of radio images. The Radio U-Net model \citep{Stuardi24} has been successfully developed and optimised for the segmentation of diffuse cluster emission, and it was recently applied to LOFAR images for the first time, demonstrating a high performance in recovering extended low-surface-brightness structures at the detection limit \citep{Stuardi25}. The application of vision transformers to the segmentation of diffuse radio emission in LOFAR images has been explored in \citet{Sanvitale25}, and other recent works applying DL to the detection of diffuse radio emission in galaxy clusters include \citet{Mishra25} and \citet{Etsebeth26}.

In this paper, we apply Radio U-Net to the third data release of the LOFAR Two-metre Sky Survey \citep[LoTSS-DR3;][]{Shimwell26}, which covers $88\%$ of the northern sky at 144 MHz. Our goal is to develop an automated DL-based pipeline capable of processing the full LoTSS-DR3 dataset and cross-matching the detected diffuse radio emission with four widely used galaxy cluster catalogues. The resulting merged catalogue provides a quantitative indicator of the presence of diffuse radio emission in each cluster, while the corresponding segmentation maps identify the location and morphology of the emission. At this stage, we do not make any attempt to automatically distinguishing different types of diffuse radio emission (e.g. radio relics, radio halos, revived fossil plasma), leaving source classification and the construction of science-ready catalogues as a task for potential users (e.g. Di Gennaro et al. in prep.). This approach is designed to provide a flexible and extensible framework for the systematic exploration of diffuse radio emission in galaxy clusters with current and future low-frequency radio surveys.

The paper is structured as follows: In Sec.~\ref{sec:catalog} we introduce the four galaxy cluster catalogues used for cross-matching and explain our strategy to create a merged catalogue. In Sec.~\ref{sec:methods} we describe survey data, the machine learning algorithm, and the detection pipeline. In Sec.~\ref{sec:results} we give an overview of the resulting catalogues and segmentation maps, while results are discussed in Sec.~\ref{sec:discussion}. We conclude and summarise our work in Sec.~\ref{sec:conclusions}. Additional materials are collected in Appendixes~\ref{app:allclusters}, \ref{app:a}, and \ref{app:examples}.

\section{A merged ICM-selected catalogue of galaxy clusters}
\label{sec:catalog}

In order to associate diffuse radio emission with individual galaxy clusters, physical information such as cluster position, redshift, and mass is required. In particular, cluster mass is a key parameter in models of diffuse radio source formation, and it is essential for statistical studies of their occurrence \citep[e.g.][]{Cassano07, Nuza12, Cuciti15}. For this work, we made use of reliable mass estimates provided by cluster catalogues based on either Sunyaev–Zeldovich detections or X-ray measurements of the thermal emission from the ICM. We used the following four catalogues:

\begin{itemize}
\item The second Planck Sunyaev–Ze'ldovich cluster catalogue \citep[PSZ2,][]{Planck16b}, with redshift updates from optical follow-ups \citep{buddendiek+15,burenin+17,burenin+18,barrena+18,streblyanska+18,Streblyanska19,AguadoBarahona19,Boada19,Zohren19,Bahk24}. This catalogue contains 1,334\footnote{The catalogue contains additional 319 clusters with no redshift information and therefore were not included in the final count.} clusters in the full sky and spans the redshift range $0.01<z<0.972$.

\item The fifth data release of the Atacama Cosmology Telescope cluster catalogue \citep[ACT-DR5;][]{hilton+20}. This catalogue spans mostly in the Southern Hemisphere ($\rm DEC<+20^\circ$, for a total of 13,168 deg$^2$) and contains 4,195 optically confirmed Sunyaev–Zeldovich-selected galaxy clusters. The catalogue is 90\% complete for masses $M_{\rm 500c}>3.8\times10^{14}~{\rm M_\odot}$, and covers a redshift range of $0.04<z<1.91$.

\item The second Meta-Catalogue of X-ray detected Clusters of galaxies, compiled from publicly available ROSAT All-Sky Survey and serendipitous X-ray cluster catalogues \citep[MCXC2;][]{sadibekova+24}, which contains 2221 clusters in the redshift range $0.003<z<1.261$. This catalogue updates the information on redshift, X-ray luminosity in the 0.2--2.4 keV range, and false detections provided by \cite{Piffaretti11} in the first release, and additionally includes cluster members from the REFLEX-II \citep{Bohringer13}, MACS \citep{Mann12,Repp18}, and RXGCC \citep{Xu22} catalogues.

\item The first SRG/eROSITA All-Sky Survey cluster catalogue \citep[1eRASS;][]{Bulbul24}. This catalogue contains 12,247 optically confirmed galaxy groups and clusters detected in the 0.2--2.3 keV range located in the western Galactic half of the sky (13,116 deg$^2$). The redshift range is $0.003<z<1.322$.
\end{itemize}

\begin{figure}
    \centering
    \includegraphics[width=0.5\textwidth]{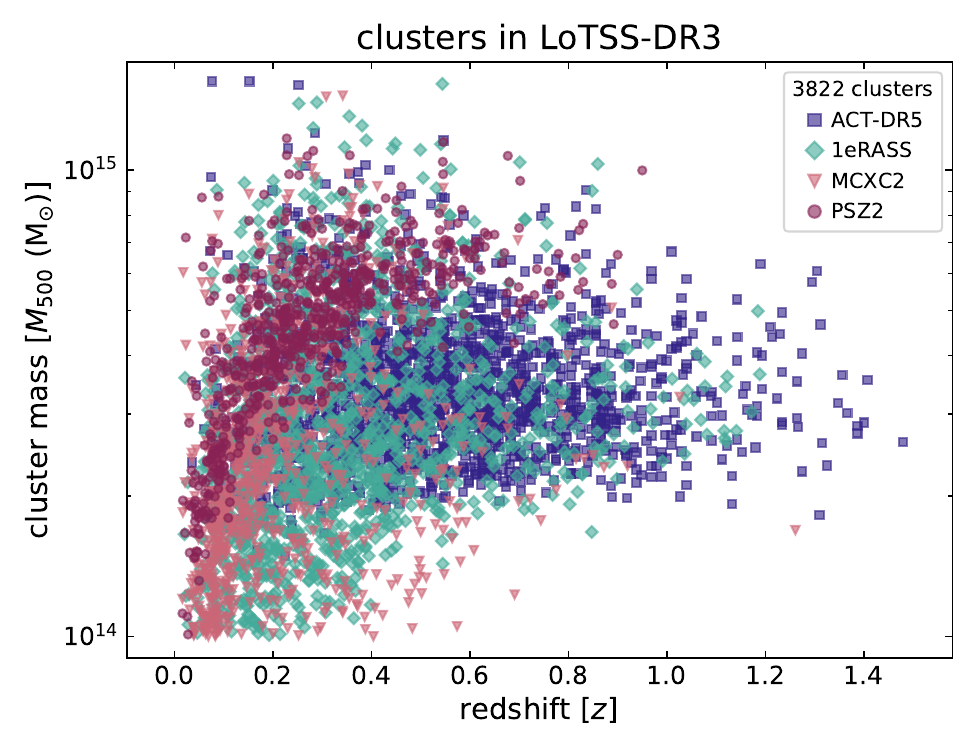}
    \vspace{-5mm}
    \caption{Mass versus redshift plot of the 3822 galaxy clusters in the LoTSS-DR3 area after the merging of the four catalogues (ACT-DR5, 1eRASS, MCXC2, PSZ2). Following our selection criteria, only clusters with a mass above $\rm 10^{14}~M_\odot$ were considered. In Appendix \ref{app:allclusters}, we show the plot of the individual catalogues before the merging.}
    \label{fig:clusters_all}
\end{figure}

   \begin{figure*}
        \centering
        \includegraphics[width=0.9\linewidth]{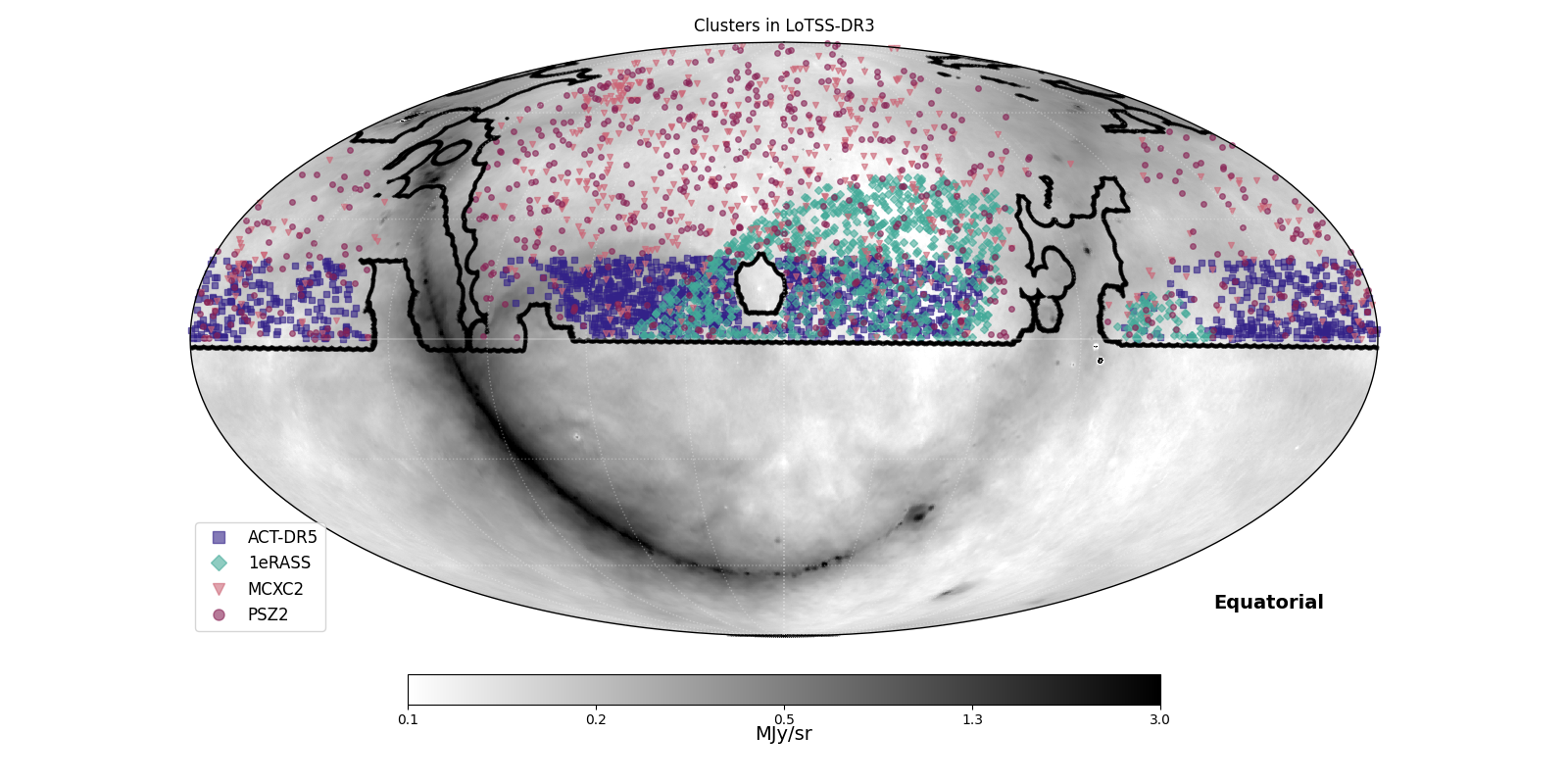}
        \caption{Position of the 3822 galaxy clusters in the merged catalogue described in Sec.~\ref{sec:catalog} overlaid on the Global Sky Model at 144 MHz derived by \citet{Zheng17}. The colour and shape of the markers represent different galaxy cluster catalogues, as explained by the legend. The black contours show the LoTSS-DR3 footprint.}
        \label{fig:clusters}
    \end{figure*}

\begin{table}
    \centering
    \caption{Galaxy clusters in the four catalogues.}
    \begin{tabular}{cccc}
    \hline
    \hline
        Catalogue & $N_{\rm LoTSS-DR3}$ & $N_{\rm merged}$ & $N_{\rm merged~\matchcal{R}}$ \\
        \hline
         PSZ2 & 686 & 686 & 626 \\
         ACT-DR5 & 1699 & 1617 & 1305 \\
         MCXC2 & 874 & 564 & 514 \\
         1eRASS & 1324 & 955 & 922 \\
         Total & 4853 & 3822 & 3367 \\
         \hline
    \end{tabular}
    \tablefoot{Column 1: Catalogue name and total count. Column 2: Number of galaxy clusters within the LoTSS-DR3 area with $M_{500}>10^{14}M_{\odot}$. Column 3: Number of galaxy clusters in the merged catalogue created as described in Sec.~\ref{sec:catalog}. Column 4: Number of galaxy clusters from the merged catalogue having an estimate of the $\mathcal{R}$ value derived from the segmentation map (i.e. observed in a mosaic of quality 0 or 1, see Sec.~\ref{fig:quality}).} 
    \label{tab:catalogues}
\end{table}

For the systems with available redshift information, we computed the corresponding $R_{500}$ starting from the original mass given in the catalogue ($M_{500}$\footnote{$M_{500}$ is defined as the total mass enclosed within a radius ($R_{500}$) where the mean density of the cluster is 500 times the critical density of the Universe at the cluster redshift.}) and assuming a standard flat $\Lambda$CDM cosmology with $H_0 = 70$ km s$^{-1}$ Mpc$^{-1}$ and $\Omega_0 = 0.3$. The same cosmology was also used to convert the physical size of $R_{500}$ into an angular scale. Duplicates across catalogues were merged when the separation of the two pairs matched within their $0.5R_{500}$ and their redshifts were consistent within the sum in quadrature of the two errors (i.e. $|z_1-z_2|<\sqrt{(\Delta z_1)^2+(\Delta z_2)^2}$). When no redshift error was provided in the catalogue, we adopted a photometric uncertainty\footnote{This is a rough estimate based on the fact that standard photometric redshift surveys report an uncertainty of $\sim0.03-0.05\times(1+z)$ for individual galaxies \citep[e.g.][]{Beck16,WH2024}, and that the cluster redshift is determined from $\sim$15–25 member galaxies, reducing the uncertainty by a factor of $\sim$4-5.} of $\Delta z = 0.01\times(1+z)$. The grouping procedure followed the order of catalogues listed above so that all PSZ2 clusters are retained in the final catalogue, while ACT-DR5 clusters are included only if not already matched to PSZ2 entries, and similarly for MCXC2 and 1eRASS. All alternative names are kept in the merged catalogue. We note that some residual duplications may still be present, due to slight discrepancies in redshift estimates or positional information among the original catalogues.

From this large ICM-selected catalogue, we selected only the systems with mass $M_{500}\geq10^{14}M_{\odot}$ that fall in the LoTSS-DR3 footprint. The final merged catalogue contains 3822 galaxy clusters, and the number of clusters belonging to each catalogue is listed in Tab.~\ref{tab:catalogues}.
Their mass-redshift distribution is shown in Fig.~\ref{fig:clusters_all} while their position is shown in Fig.~\ref{fig:clusters}, overlaid on the Global Sky Model at 144 MHz derived by \citet{Zheng17} with PyGDSM\footnote{\url{https://github.com/telegraphic/pygdsm?tab=readme-ov-file}}. The four catalogues are not uniformly represented in the sample and are also unevenly distributed in the LoTSS-DR3 sky area, with the largest ACT-DR5 sample covering only the low-declination region by construction of the ACT survey.

\section{Methods}
\label{sec:methods}

In this section, we describe the methods that we have used to provide information on the presence of diffuse radio emission for each cluster in the merged catalogue created in Sec.~\ref{sec:catalog}. In particular, we first describe the LoTSS-DR3 (Sec.~\ref{sec:lotss}) and the main characteristics of the Radio U-Net architecture (Sec.~\ref{sec:radiounet}), together with its implementation to process LoTSS-DR3 data and quality assessment of resulting segmentation maps (Sec.~\ref{sec:quality}). In Sec.~\ref{sec:detpipeline}, we introduce the detection pipeline implemented to generate the final catalogue starting from the merged catalogue of galaxy clusters.

\subsection{The LOFAR Two-metre Sky Survey third data release}
\label{sec:lotss}

The LoTSS is a wide-area 120-168\,MHz survey aiming to cover the entire northern sky. Three partial LoTSS data releases have been published in the past years: a preliminary data release \citep{Shimwell17}, and two full-quality releases, DR1 and DR2 \citep{Shimwell19, Shimwell22}. These releases provide Stokes $I$, $Q$, $U$, and $V$ images, associated catalogues, and calibrated uv-data. The DR2 covered approximately 27\% of the northern sky.

The recent DR3 \citep{Shimwell26} extends the sky coverage to 88\% of the northern hemisphere, collecting 12,950 hours of observations over a period of 10.5 years. The final 120--168\,MHz continuum images have angular resolution of 6$\arcsec$ above declination $10^{\circ}$, increasing to 9$\arcsec$ between $0^{\circ}$ and $10^{\circ}$. The median rms sensitivity of the mosaics is $92\,\mu$Jy\,beam$^{-1}$, improving to $68\,\mu$Jy\,beam$^{-1}$ at high elevation but degrading to $183\,\mu$Jy\,beam$^{-1}$ near the celestial equator, primarily due to projection effects that reduce the effective station collecting area. Low-resolution images with a common 20$\arcsec$ resolution across the whole area are also released.

Upon completion, LoTSS will consist of 3168 individual pointings. The LoTSS-DR3 includes all fields observed before the suspension of LOFAR operations in September 2024 for the LOFAR~2.0 upgrade, excluding most of those within $10^{\circ}$ of extremely bright radio sources (Cassiopeia~A, Cygnus~A, Virgo~A, Taurus~A, and Hercules~A) which produce severe contamination in their surrounding fields. In total, 2551 pointings are included in DR3.

In many cases, the LoTSS-DR2 images are directly reused in DR3 ($85\%$ of DR2 pointings), whereas in others the images have been reprocessed using updated calibration pipelines or extended with additional observations. As with DR2, mosaics are constructed from a weighted combination of all datasets contributing to a given sky position, with the weighting determined by the local rms noise and the LOFAR primary beam response. Although DR3 adopts a revised mosaicking strategy designed to optimise the combination of overlapping observations, in this work, we used mosaics created with the same strategy as in DR2 (i.e. having the same size and name as individual LoTSS pointings) since the analysis was developed before the implementation of the new approach. Since mosaics can be formed at any position on the sky, this choice does not affect the quality of the data.

While LOFAR is well suited to detect extended and diffuse radio emission, thanks to its excellent uv-coverage at the shortest baselines, the automated calibration and imaging strategy adopted for the LoTSS is not optimised for such emission. As outlined by \citet{Shimwell22}, it is unavoidable that extended low-surface-brightness emission is occasionally not recovered by the deconvolution algorithm during calibration, leading to a partial suppression of its flux if the unrecovered flux is significant compared to the total in the field. Furthermore, if the emission is not completely deconvolved in the final image, its apparent brightness can be different from the real one \citep[see section 3.4 of][for a quantification of this effect]{Shimwell22}. The latter effect is mitigated in low-resolution images at $20\arcsec$, where diffuse radio emission is generally deconvolved. 

When measuring the flux densities of diffuse radio sources, dedicated pipelines are required for accurate source extraction and self-calibration of individual targets \citep{vanWeeren21}. This approach was adopted, for example, for the sample of 309 galaxy clusters from the PSZ2 catalogue within the LoTSS-DR2 area (hereafter, the LoTSS-DR2/PSZ2 sample), presented by \citet{Botteon22}. Applying such a procedure to a single galaxy cluster observed in four pointings may require 24–36 hours on a machine with 512 GB of RAM and 64 AMD cores, with the total processing time scaling linearly with the number of clusters. Consequently, the selection of the most suitable targets in LoTSS-DR3 for further processing with this dedicated pipeline -- which will be the subject of future works -- represents a key motivation for the present study.

\subsection{Radio U-Net}
\label{sec:radiounet}

Radio U-Net\footnote{https://github.com/ICSC-Spoke3/Radio-U-Net} is a convolutional neural network developed to identify diffuse radio emission in interferometric images \citep{Stuardi24}. The network has been implemented using Python and exploiting the KERAS software package \citep{Chollet15}, distributed as part of the Tensorflow framework \citep{Abadi15}, version 2.3.0. Here we only give an overview of the network design and implementation. We refer the reader to \citet{Stuardi24} for a full description of the network architecture, training dataset, and procedure as well as validation tests.

The model follows the U-Net design \citep{Ronneberger15}, consisting of a contracting path and a symmetric expanding path. In the contracting path, the input image is progressively compressed through successive convolution, activation, normalisation, and pooling layers, allowing the network to learn increasingly abstract features of the image. In particular, in our implementation, at each of the four contracting levels two convolutional layers are used for feature extraction, followed by a ReLu (Rectified Linear Unit) activation function to introduce non-linearity \citep{Agarap18}. Batch normalisation is used after each convolutional layer to improve convergence, and the dimensionality reduction is obtained with a max pooling function. The expanding path reconstructs the image to its original resolution in four levels, using transpose convolutions, followed by ReLu activation, and upsampling layers. 

The final softmax layer produces pixel-by-pixel information about the presence of the diffuse radio emission. The output image has the same size as the input image, but pixel values range from 0 to 1, with higher values indicating the presence of diffuse radio emission. No thresholding is applied by the network to the segmented maps to preserve the full probabilistic information. However, a threshold can be introduced to create a binary map for visualisation and qualitative assessment purposes (e.g. in Sec.~\ref{sec:quality}). This type of pixel-level classification is named semantic segmentation, and it is particularly effective for the detection of complex and multi-scale objects, providing information on both their shape and position in large images \citep[see][for a review]{Csurka22}. 

The network was trained using synthetic radio images derived from cosmological magneto-hydrodynamical simulations, where diffuse radio emission was modelled from shock-accelerated electrons and projected into mock lightcones \citep{Vazza19}. More details about the creation of these synthetic sky images are given in \citet{Gheller22}, section 4 of \citet{Stuardi24}, and references therein. These simulations do not include radio emission generated by turbulent re-acceleration, as expected for radio halos or bridges \citep{Brunetti11b, Brunetti20}. Nevertheless, the resulting images reproduce a range of morphologies and surface brightness distributions that overlap with those observed in radio halos and bridges, and previous works have shown that such sources can still be effectively detected by our approach \citep{Stuardi24, Sanvitale25}. 

Synthetic sky images of diffuse radio emission were then processed to mimic LOFAR HBA observations at $6\arcsec$ resolution, including Gaussian noise and imperfect deconvolution, so as to reproduce typical observational artefacts. They do not include other real radio sources. These mock images were used as input to the network, while a binary reference masks derived from the underlying sky images were employed to train the network to identify regions containing diffuse radio emission above $10^{-8}$ Jy pixel$^{-1}$. This value is a factor $\sim10^{-3}$ lower than the median noise level to train
the network to recognise emission structures in low signal-to-noise conditions. Input images are also transformed in logarithmic scale and normalised between 0 and 1, using $10^{-10}$ and $10^{-5}$ Jy pixel$^{-1}$ as minimum and maximum data normalisation values. Pixels outside this range were clipped to the boundaries. The use of min–max normalisation does not lead to a significant loss of information, because it is applied after a logarithmic transformation of the input images, which enhances low-surface-brightness emission, while normalisation bounds are chosen to encompass the typical range relevant for diffuse radio emission.  Synthetic sky models and images used for training are publicly available \footnote{\url{https://owncloud.ia2.inaf.it/index.php/s/IbFPlCCcPUresrr}}.

To handle large input sizes (2000$\times$2000 pixels), the images were divided into 192$\times$192 pixel tiles. The training set comprises 100 images (i.e. 10000 tiles), but a fraction of them (5$\%$, that is, 500 tiles) is kept as a validation set to monitor the performance of the network during training. This choice represents a trade-off between ensuring a reliable estimate of the model performance and maximising the amount of data available for training, which is crucial for DL models. The tiles are all independently processed during training and validation. During the inference phase (i.e. when the network is applied to new images), tiles are instead extracted with partial overlap. Although each tile is still processed independently, only the central regions are retained and subsequently reassembled, in order to minimise boundary effects and avoid discontinuities at tile edges.

Training relied on categorical cross-entropy as the loss function and RMSprop as the optimiser \citep{Tieleman12,Goodfellow16}. Categorical cross-entropy measures how well the predicted probabilities match the true class (0 or 1 in our case), strongly penalising confident wrong predictions. The loss function is minimised during the training to obtain accurate predictions. Hyperparameters (e.g. learning rate $10^{-4}$, batch size 50, 200 epochs) were selected following earlier optimisation, where the training was monitored using the validation loss to assess convergence and prevent overfitting \citep{Stuardi24}.

Once trained, the model can process full-size simulated images (i.e. 2000$\times$2000 pixels) in less than a second on a single NVIDIA Tensor Core GPU. This efficiency makes Radio U-Net particularly well-suited for large-scale surveys. In \citet{Stuardi24}, the network was applied to archival LoTSS-DR2 cut-outs of a test sample of galaxy clusters. The input images were first converted to a logarithmic scale and normalised to the range $10^{-7}$–$10^{-2}$ Jy beam$^{-1}$. With respect to previous work, only the data ingestion stage of the pipeline (responsible for reading the input images) was modified, in order to handle images of varying sizes and to process LoTSS mosaics with non-uniform fields of view.

The entire LoTSS-DR3 at $20\arcsec$ resolution, for a total of 2551 mosaics of size about 3000$\times$3000 pixels, was processed with Radio U-Net in four hours, by distributing parallel jobs on the Leonardo\footnote{\url{https:// www.hpc.cineca.it/ systems/ hardware/ leonardo/}} supercomputer at the CINECA Italian Supercomputing Centre and offloading data reading into GPUs. The resulting dataset, consisting of thousands of FITS files matching the size of the original LoTSS pointings, occupies 110 GB.

\subsection{Quality assessment of segmented images}
\label{sec:quality}

\begin{figure*}
        \centering
        \includegraphics[width=\linewidth]{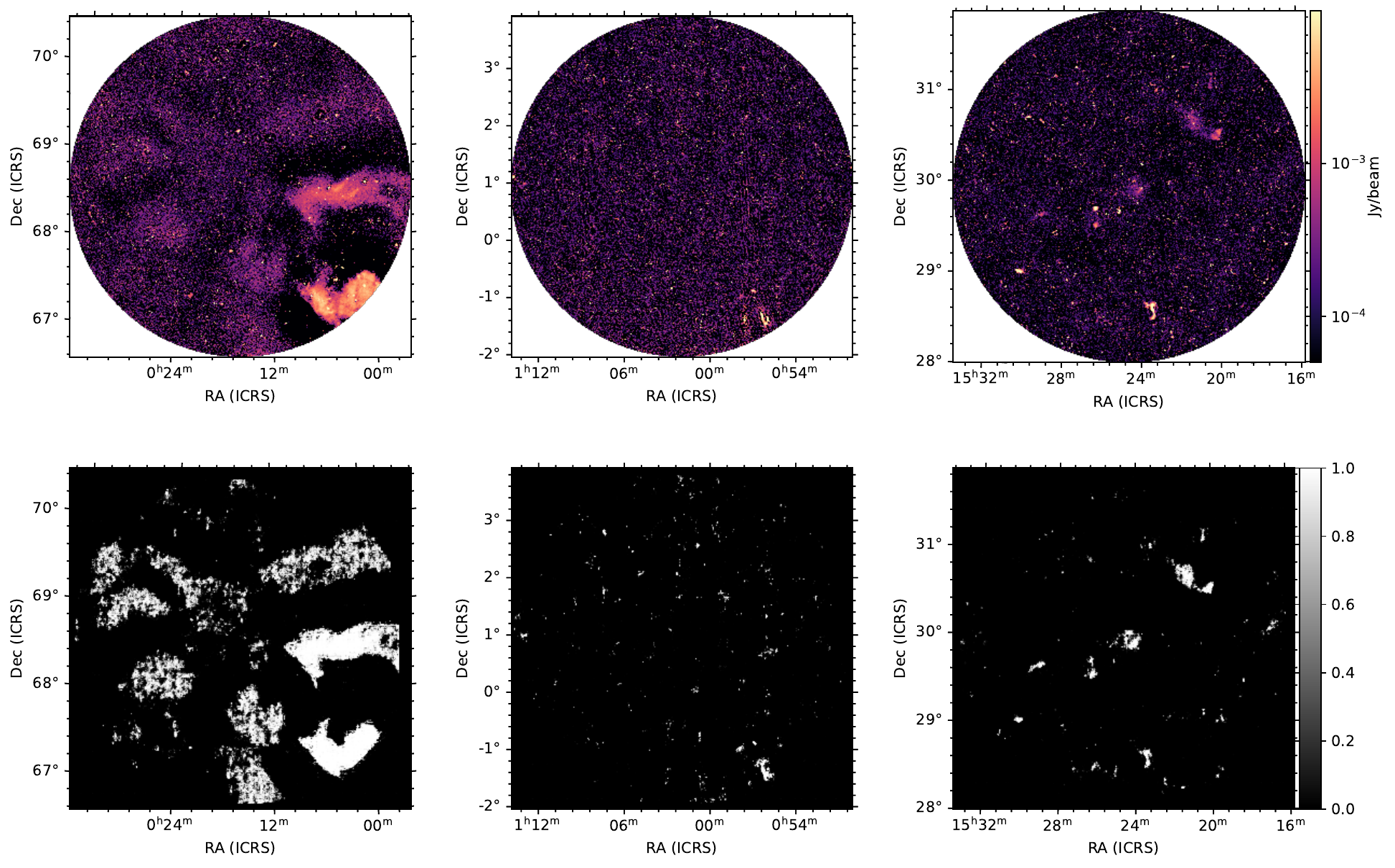}
        \caption{Representative example of quality 2 (top and bottom left panels), quality 1 (top and bottom central panels), and quality 0 (top and bottom right panels) mosaics. The top row shows the LoTSS-DR3 mosaics with a common colour-scale. The bottom row shows the corresponding segmented maps produced by Radio U-Net. While quality class~2 mosaics show clear large-scale patterns, quality class~1 mosaics are characterised by a scattered distribution of erroneously segmented pixels.}
        \label{fig:pointings}
    \end{figure*}

From an initial visual inspection of the segmented images, we noticed that some pointings were incorrectly segmented because of the presence of Galactic diffuse radio emission. This effect occurred mostly at Galactic latitude $|l|<15^\circ$ but even up to $l=75^\circ$ in the region of the North Polar Spur. An example of a pointing affected by this problem is shown in Fig.\ref{fig:pointings}, left panel, with the corresponding segmentation map in the bottom panel. Galactic diffuse radio emission, which is extended over large angular scales and is observed in the form of large-scale ripples in the LoTSS survey, is recognised and segmented by the network. Galactic emission was not present in the original training dataset, which was designed to reproduce LoTSS-DR2 data quality and tested on sky regions where the Galactic foreground is negligible. As a consequence, the network is not able to exclude this emission from the segmentation maps, with the risk of misidentifying genuine diffuse radio emission of extragalactic origin.

In addition, we found that other pointings — mostly at low declination ($\delta<15^\circ$) and at Galactic latitude $15^\circ<|l|<30^\circ$ — show some signs of noisy segmentation, such as an increased number of scattered pixels with value approaching unity. A representative example is shown in Fig.~\ref{fig:pointings}, central panel, with the corresponding segmentation map in the bottom panel. This is likely due once again to the fact that the network cannot perfectly generalise to sky regions not represented in the training set. Still, in these pointings, the network correctly detects known diffuse radio emission of extragalactic origin. We therefore decided to retain these intermediate quality pointings in our analysis and provide segmentation maps for them, but with a distinct quality classification.

We note that segmentation quality is not directly related to the noise level of the mosaics, since large-scale Galactic ripples typically do not significantly affect the rms estimate, which is computed on small scales. Moreover, there is no exact correspondence between the presence of Galactic emission in LoTSS images and the surface brightness of the Galaxy derived from single-dish maps since the minimum baseline in regular LoTSS processing is 100 m. These considerations motivate a visual classification rather than cuts based on image noise or Galactic emission. The classification was performed on segmentation maps clipped at a threshold of 0.5, chosen as an intermediate value between 0 and 1 for visualisation purposes only.

Three classes were defined as follows:

\begin{itemize}
\item Class 0: Mosaics with segmented images of the same quality as LoTSS DR2 mosaics, where Radio U-Net was originally assessed.
\item Class 1: Mosaics with correctly segmented images but generally showing a higher noise level (i.e. $>150$ $\mu$Jy beam$^{-1}$), resulting in an increased number of scattered pixels with a value larger than 0.5. While quality class~1 mosaics can still be used to search for diffuse radio emission, we recommend exercising caution when inspecting these segmented maps, as they are more prone to misidentifying imaging artefacts as diffuse structures.
\item Class 2: Mosaics with clear segmentation errors easily visible as coherent stripes of pixels with values larger than 0.5. In order to avoid contamination of genuine diffuse cluster emission by Galactic foregrounds, we decided to exclude these pointings from further analysis.
\end{itemize}

\begin{figure*}
        \centering
        \includegraphics[width=0.9\linewidth]{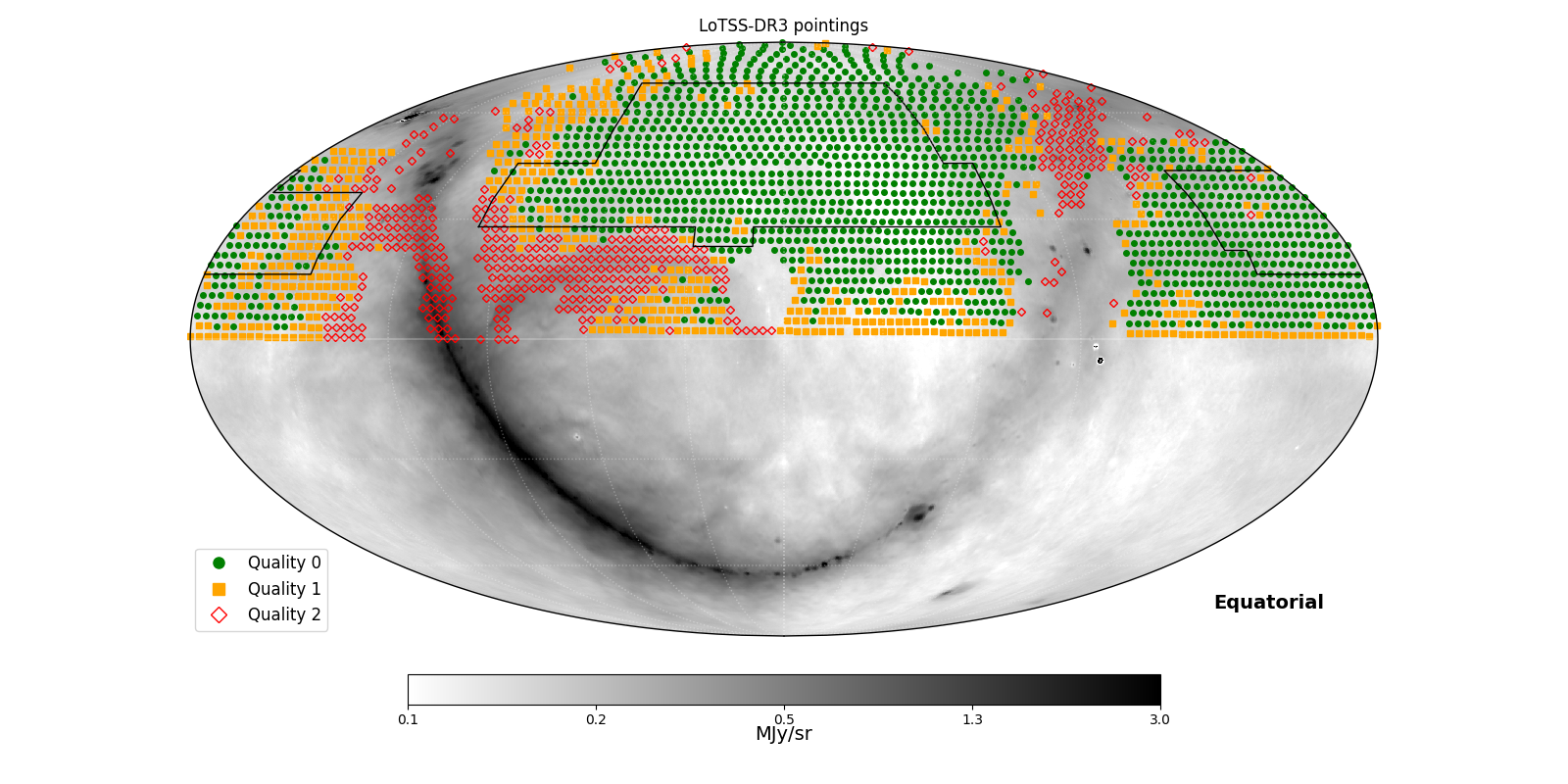}
        \caption{Position of the 2551 pointings of the LoTSS-DR3 overlaid on the Global Sky Model at 144 MHz derived by \citet{Zheng17}. The colour and shape of the markers represent the quality class given in Sec.~\ref{sec:quality}, as explained by the legend. In particular, red empty diamonds are quality~2 pointings, which were excluded by this analysis mainly due to the contamination of the Milky Way emission. Black lines show the LoTSS-DR2 footprint.}
        \label{fig:quality}
    \end{figure*}

Among a total of 2551 LoTSS-DR3 mosaics, 1545 ($60\%$) are of quality 0, 579 ($23\%$) of quality 1, and 427 ($17\%$) of quality 2. The position of the mosaics coloured by the three quality classes is shown in Fig.\ref{fig:quality}, overlaid on the Global Sky Model at 144 MHz. Their distribution is a direct consequence of the fact that we are extrapolating Radio U-Net results to sky regions characterised by substantially higher noise levels, and the network exhibits limited generalisation capability under these conditions. A possible solution to this problem is discussed in Sec.~\ref{sec:discuss_ML}, but its implementation is out of the scope of this work.

We note that excluding the Galactic region (i.e. quality~2 mosaics) from the subsequent analysis does not significantly affect the detection pipeline based on the merged catalogue of galaxy clusters, as these catalogues already avoid most of it (see Fig.~\ref{fig:clusters}). In total, 455 out of 3822 galaxy clusters in the merged catalogue lie in regions classified with quality 2, of which 312 belong to the ACT-DR5 catalogue (a summary is in Tab.~\ref{tab:catalogues}).

\subsection{The detection pipeline}
\label{sec:detpipeline}

While the segmented images produced by Radio U-Net can be directly exploited for blind searches of diffuse radio emission across the entire LoTSS-DR3, additional physical information is required to assess whether the detected emission is associated with a galaxy cluster. In particular, cluster position, redshift, and mass are needed and can be retrieved from the merged catalogue created in Sec.~\ref{sec:catalog}. Excluding the 455 galaxy clusters lying in quality~2 mosaics, we applied the detection pipeline on the remaining 3367 galaxy clusters. The number of galaxy clusters considered in the subsequent analysis is reported in Tab.~\ref{tab:catalogues}, subdivided among the four original catalogues ($N_{\rm merged~\matchcal{R}}$).

Using the position and angular size of each of the 3367 galaxy clusters, we computed for every system a single quantity that traces the presence of diffuse radio emission within the cluster’s projected area. Following \citet{Stuardi24}, we defined the $\mathcal{R}$ value as the ratio between the sum of pixel values in the segmented image within a circle of radius $2.2R_{500}$ centred on the cluster and the total number of pixels enclosed by that circle ($N_{2.2R_{500}}$):

\begin{equation}
    \mathcal{R} = \frac{\sum^{2.2R_{500}}_{m,n} s_{m,n}}{N_{2.2R_{500}}},
\end{equation}
where $s_{m,n}$ is the output of the network at the $(m,n)$ pixel. By design, $\mathcal{R}$ approaches zero when no source is detected and reaches one when a diffuse radio emission completely fills the circle. The value of 2.2$R_{500}$ was chosen, as it is the maximum distance at which radio relics were detected in the LoTSS-DR2/PSZ2 sample \citep{Jones23}. This makes $\mathcal{R}$ an effective discriminator between clusters with and without diffuse radio emission. 

In \citet{Stuardi24}, using the galaxy cluster catalogue presented by \citet{Botteon22} as a benchmark, we showed that more than $80\%$ of PSZ2 clusters in LoTSS-DR2 hosting diffuse radio emission have $\mathcal{R} > 0.015$. Adopting this threshold to separate clusters with diffuse emission (DE) from those without (NDE) results in an overall classification accuracy of $73\%$, defined as the fraction of correctly classified clusters over the total sample. We note that galaxy clusters hosting radio sources classified as `uncertain' by \citet{Botteon22} -- that is, sources whose emission is significantly affected by imaging artefacts or whose morphology, size, and/or position are inconsistent with standard halo or relic classifications -- are included in the DE category. Although not clearly classified, these systems do host diffuse radio emission, making the $\mathcal{R}$ value sensitive also to revived fossil plasma sources or radio phoenixes \citep[see][for a recent work on these sources]{Bruno25}.

For each cluster in the merged catalogue, we annotated the survey pointings in which it was observed, its angular distance from the pointing centre, and the local noise at the cluster position in each individual pointing prior to mosaicking. This information is useful for any subsequent reprocessing of individual targets, where a careful selection of the relevant pointings may be required. It is also important in light of the different mosaicking strategies adopted in the LoTSS DR2 and DR3. In the DR2 scheme \citep{Shimwell22}, mosaics retain the same footprint as individual pointings and may therefore overlap. In contrast, DR3 mosaics are generated on a HEALPix grid and do not overlap. Although a given cluster should appear identical in multiple mosaics, the segmentation quality of the mosaics can vary, since each mosaic covers a different region of the sky.

When a cluster was observed in multiple pointings, the $\mathcal{R}$ value was measured in the quality~0 mosaic where the cluster lies closest to the centre. If no quality~0 mosaic was available or if the cluster (within its 2.2$R_{500}$) fell outside the mosaic, the nearest quality~1 mosaic was used. This quality is stored as a quality class for the $\mathcal{R}$ value (0 or 1). Of the 3367 clusters outside of quality~2 regions, 2216 have quality class~0 and 1151 have quality class~1. In rare cases where a cluster lies at the boundary between two mosaics, the segmented mosaics were first merged before estimating $\mathcal{R}$. This occurred only seven times, and for these, the mosaics combined to compute $\mathcal{R}$ are listed in the `Comment' column of the catalogue. We also annotated cases in which the $\mathcal{R}$ value was computed from a mosaic whose corresponding pointing is not present in the LoTSS-DR3 but instead includes only the edge regions of neighbouring pointings. This situation arises from the use of the DR2 mosaicking strategy and affects only 28 clusters located near the edges of the DR3 footprint. Furthermore, in the `Overlap' column, we list any clusters whose projected positions overlap within $2.2R_{500}$ on the sky plane, even when they lie at different redshifts. Since $\mathcal{R}$ does not account for overlaps, the computed value integrates over all overlapping systems.

Statistical uncertainties on the $\mathcal{R}$ values were derived on the basis of the results obtained in \citet{Stuardi24}, as detailed in Appendix~\ref{app:a}. Uncertainties increase with the redshift of the galaxy cluster. To define a threshold in $\mathcal{R}$ for distinguishing DE from NDE clusters, we followed \citet{Stuardi24}, computing classical evaluation metrics (accuracy, precision and recall) for a test sample of 246 LoTSS-DR2/PSZ2 galaxy clusters from \citet{Botteon22} as functions of the chosen $\mathcal{R}$ value. Unlike the previous analysis, we now include the uncertainty in $\mathcal{R}$ for each cluster.

Each cluster $i$ in a sample of $N$ galaxy clusters is associated with an observed value $\mathcal{R}_i$ and measurement uncertainty $\sigma_i$.  
We assumed the true value, $\mathcal{R}_i^{\mathrm{true}}$, follows a normal distribution. For a classification threshold, $\mathcal{R}_t$, the probability that object $i$ belongs to class DE  
(i.e. that its true value exceeds the threshold) is
\[
p_i(\mathcal{R}_t) = P(\mathcal{R}_i^{\mathrm{true}} > \mathcal{R}_t)
       = 1 - \Phi\!\left(\frac{\mathcal{R}_t - \mathcal{R}_i}{\sigma_i}\right),
\]
where $\Phi$ is the cumulative distribution function of the standard normal distribution.

Let $y_i \in \{0,1\}$ be the true class labels of the 246 galaxy clusters (NDE and DE) chosen as a test sample from \citet{Botteon22}.  
The expected counts of true positives (TP), false positives (FP),  
true negatives (TN), and false negatives (FN) for threshold $\mathcal{R}_t$ are
\begin{align*}
\mathbb{E}[\mathrm{TP}(\mathcal{R}_t)] &= \sum_i y_i\,p_i(\mathcal{R}_t), \\
\mathbb{E}[\mathrm{FP}(\mathcal{R}_t)] &= \sum_i (1-y_i)\,p_i(\mathcal{R}_t), \\
\mathbb{E}[\mathrm{TN}(\mathcal{R}_t)] &= \sum_i (1-y_i)\,(1-p_i(\mathcal{R}_t)), \\
\mathbb{E}[\mathrm{FN}(\mathcal{R}_t)] &= \sum_i y_i\,(1-p_i(\mathcal{R}_t)).
\end{align*}

From these, we computed the expected accuracy (fraction of correct classifications to the total number of objects), precision (the ratio of the TP to the total
number of objects identified as positive), and recall (the ratio of the TP to the total number of DE objects):
\begin{align*}
\mathbb{E}[\mathrm{Accuracy}(\mathcal{R}_t)] &= 
    \frac{\mathbb{E}[\mathrm{TP}(\mathcal{R}_t)] + \mathbb{E}[\mathrm{TN}(\mathcal{R}_t)]}{N}, \\
\mathbb{E}[\mathrm{Precision}(\mathcal{R}_t)] &= 
    \frac{\mathbb{E}[\mathrm{TP}(\mathcal{R}_t)]}{\mathbb{E}[\mathrm{TP}(\mathcal{R}_t)] + \mathbb{E}[\mathrm{FP}(\mathcal{R}_t)]}, \\
\mathbb{E}[\mathrm{Recall}(\mathcal{R}_t)] &= 
    \frac{\mathbb{E}[\mathrm{TP}(\mathcal{R}_t)]}{\sum_i y_i}.
\end{align*}
While precision highlights the purity of DE predictions, the recall focuses on the network’s ability to identify all DE galaxy clusters. These definitions are the same used by \citet{Stuardi24}, with the only variation of considering the uncertainty on the $\mathcal{R}$ value in the computation.

   \begin{figure}
        \centering
        \includegraphics[width=\linewidth]{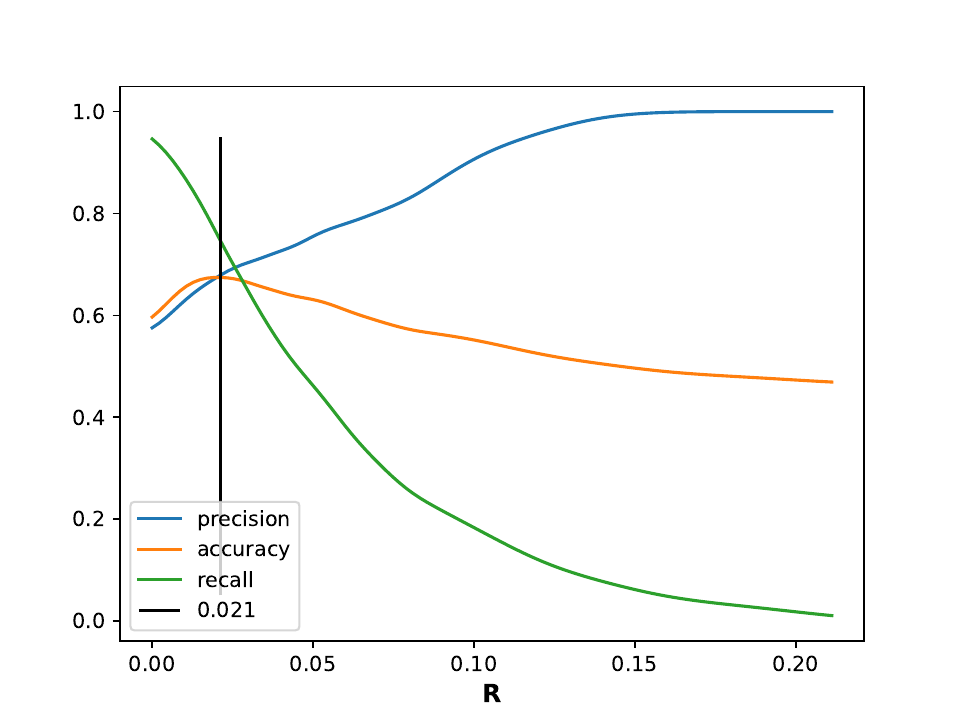}
        \caption{Binary classification metrics for the test sample of 246 LoTSS-DR2/PSZ2 galaxy clusters \citep{Botteon22} as a function of the value of $\mathcal{R}$ used as a threshold, considering its uncertainties.}
        \label{fig:pr_curve}
    \end{figure}

The optimal threshold $\mathcal{R}_t^\ast$ is chosen as the value of $\mathcal{R}_t$ that maximises the selected performance metric. Once $\mathcal{R}_t^\ast$ is determined, clusters are classified deterministically: as DE if $\mathcal{R}_i > \mathcal{R}_t^\ast$ and as NDE otherwise. As shown in Fig.~\ref{fig:pr_curve}, the classification accuracy peaks at $\mathcal{R}_t^\ast=0.021$ where the accuracy is $67\%$, the precision is $68\%$, and the recall is $74\%$. Choosing a higher precision threshold helps build a sample with fewer false detections of DE, whereas prioritising higher recall reduces the risk of missing DE galaxy clusters, at the cost of increasing FPs.

To highlight potentially uncertain classifications, it is possible to introduce a binary classification flag.  
A classification is considered reliable if its measured $\mathcal{R}$ value lies more than one standard deviation from the threshold:
\begin{equation}
    |\mathcal{R}_i - \mathcal{R}_t^\ast| > \sigma_i.
\end{equation}
This approach ensures that both threshold selection and individual measurement uncertainties are explicitly accounted for in the final classification. 

In the case that $\mathcal{R}_t^\ast=0.021$ is chosen as a threshold, we verified that classification metrics improve using the same test sample from \citet{Botteon22} but excluding clusters with an unreliable classification flag. Although this further selection reduced the test sample to 174 clusters, it allows us to achieve $76\%$ accuracy, $74\%$ precision, and $85\%$ recall in the classification—meaning that $85\%$ of clusters hosting genuine diffuse radio emission are correctly identified as DE. Out of the correctly identified DE galaxy clusters, 54 host radio halos or candidate radio halos, 22 host at least a radio relic or a candidate, and 18 host only a diffuse radio emission classified as uncertain by \citet{Botteon22}. 

\begin{figure}
        \centering
        \includegraphics[width=0.8\linewidth]{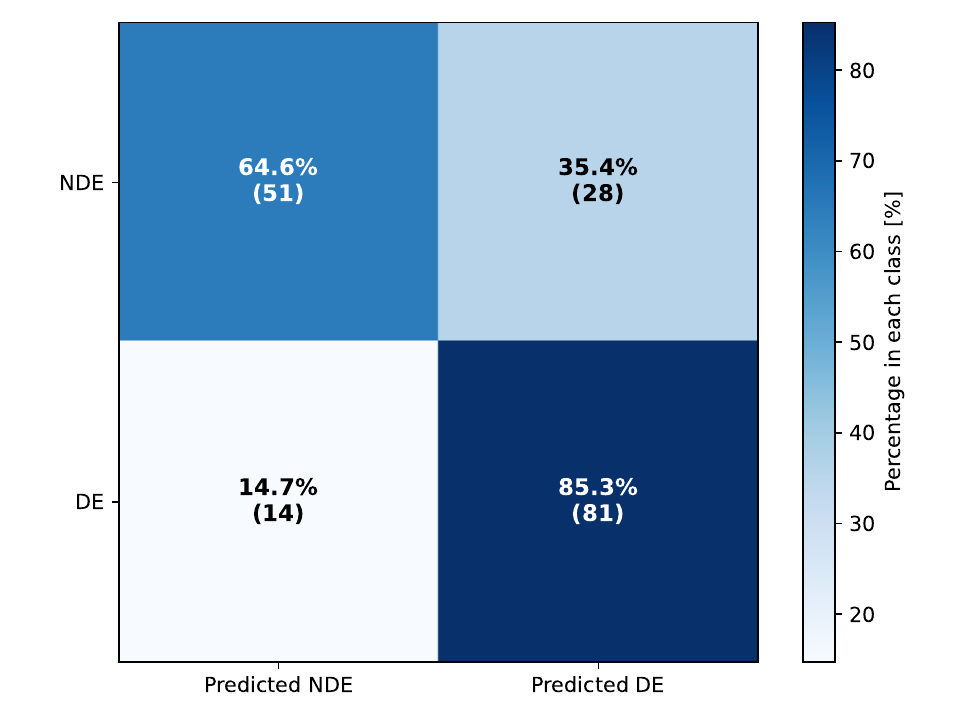}
        \caption{Confusion matrix obtained for the test sample of 174 galaxy clusters with a good classification flag using $\mathcal{R}_t^*=0.021$ as a detection threshold. The number in brackets is the total number in each box.}
        \label{fig:confmatrix}
    \end{figure}

The confusion matrix obtained with the detection pipeline on this test sample of 174 galaxy clusters is shown in Fig.~\ref{fig:confmatrix}. Notice, however, that this residual sample is slightly biased, counting for 95 DE clusters and 79 NDE clusters. We note that a relatively large fraction of NDE systems is classified as DE by the network ($35\%$). This may have two main causes. First, some NDE systems may host extended radio emission associated with radio galaxies, which the network identifies as DE. Second, the NDE classification based on visual inspection may be uncertain in borderline cases, where human judgement is also intrinsically limited, and faint diffuse radio emission may be overlooked. In some cases, the network may, in fact, detect low brightness emission that is close to the noise level and difficult to identify.

The detection pipeline thus provides for each galaxy cluster a quantitative indicator -- that is, the $\mathcal{R}$ value -- which can be used, together with its uncertainty, to identify clusters hosting diffuse radio emission. The choice of $\mathcal{R}_t^\ast=0.021$ as a threshold to separate clusters with and without diffuse radio emission maximises the accuracy of this classification. However, since the above performance metrics are computed on the LoTSS-DR2/PSZ2 galaxy cluster sample, which is mostly located in high-quality LoTSS regions, we note that, although for clusters with quality class~1 the $\mathcal{R}$ value still provide an indication of the presence of diffuse radio emission, the classification performance based on the $\mathcal{R}$ value may be less reliable. Furthermore, it is not possible to verify how the classification metrics depend on the original galaxy cluster catalogue since, to date, the only reference sample is based on the PSZ2 catalogue.

\section{Results}
\label{sec:results}

In this section, we first describe the main results of this work (Sec.~\ref{sec:results1}), which consist of the following:

\begin{itemize}
\item Segmentation maps of the LoTSS-DR3 produced with Radio U-Net, associated with the full list of 2551 LoTSS-DR3 pointings and the quality classes associate to each mosaic (having the same shape and name of individual pointings as for the DR2 mosaicking scheme), as described in Sec.~\ref{sec:quality};
\item A merged catalogue of 3822 galaxy clusters within the LoTSS-DR3 footprint, of which 3367 have an estimated $\mathcal{R}$ value obtained from Radio U-Net results.
\end{itemize}
Finally, we illustrate how these results can be used to select a sub-sample of galaxy clusters with a high probability of hosting diffuse radio emission and provide some examples (Sec.~\ref{sec:goldsample}).

\subsection{Segmentation maps and a merged catalogue of galaxy clusters with $\mathcal{R}$ values}
\label{sec:results1} 

Example of segmentation maps produced by Radio U-Net are shown in Fig.~\ref{fig:pointings}. These maps provide pixel-level information about the presence of diffuse radio emission as recognised by the network. The resolution of these images is the same as the original LoTSS images at $20\arcsec$, having a pixel size of $4.5\arcsec$. Values approaching 1 are recognised by the network as part of diffuse radio emission. Generally, the boundaries of these regions are very sharp, with values decreasing below 0.5 within 4-5 pixels.

As extensively discussed in \citet{Stuardi24}, Radio U-Net tends to misidentify radio emission from extended, low–surface-brightness active galactic nuclei (AGNs) as diffuse radio emission related to the ICM. This behaviour arises because AGN-related emission was not included in the original training set. A similar misidentification can also occur for nearby galaxies, whose radio emission may appear approximately circular and thus resemble that of radio halos. Finally, as discussed in Sec.~\ref{sec:quality}, pointings characterised by the presence of Galactic diffuse radio emission or by higher noise levels are more prone to misidentification by the network (see two examples in Fig.~\ref{fig:pointings}).

To provide a standardised access and visualisation of the segmentation maps, we have created a Hierarchical Progressive Surveys \citep[HiPS,][]{Fernique15} map of all quality classes~1 and~0 mosaics, displaying only pixels with values above 0.9. This threshold was chosen to optimise the visualisation of large fields of view, highlighting only regions detected with the highest probability. This map is available on the dedicated page of the survey website\footnote{\url{https://lofar-surveys.org/clusters_dr3.html}} and can be visualised on Aladin\footnote{\url{https://aladin.cds.unistra.fr/hips/}} for the inspection of the full DR3 area (excluding Galactic regions). We also provide a table listing the 2551 LoTSS-DR3 pointings with the associated mosaic's quality class, while segmentation maps of single mosaics are made available upon request.

While segmentation maps can be used to readily identify the presence of diffuse radio emission, a comparison with multi-wavelength data is required to associate the emission detected by the network with individual objects. For this reason, and building on the results presented by \citet{Stuardi24}, we cross-matched the segmentation maps produced by Radio U-Net with the position of galaxy clusters contained in the merged catalogue created in Sec.~\ref{sec:catalog}. Therefore, as explained in Sec.~\ref{sec:detpipeline}, for each of the 3822 galaxy clusters in our merged catalogue, we found the reference pointing and, for those in quality class~0 and 1 pointings, we computed the $\mathcal{R}$ value from the segmentation map. The merged catalogue thus provides the following information:

   \begin{figure}
        \centering
        \includegraphics[width=0.9\linewidth]{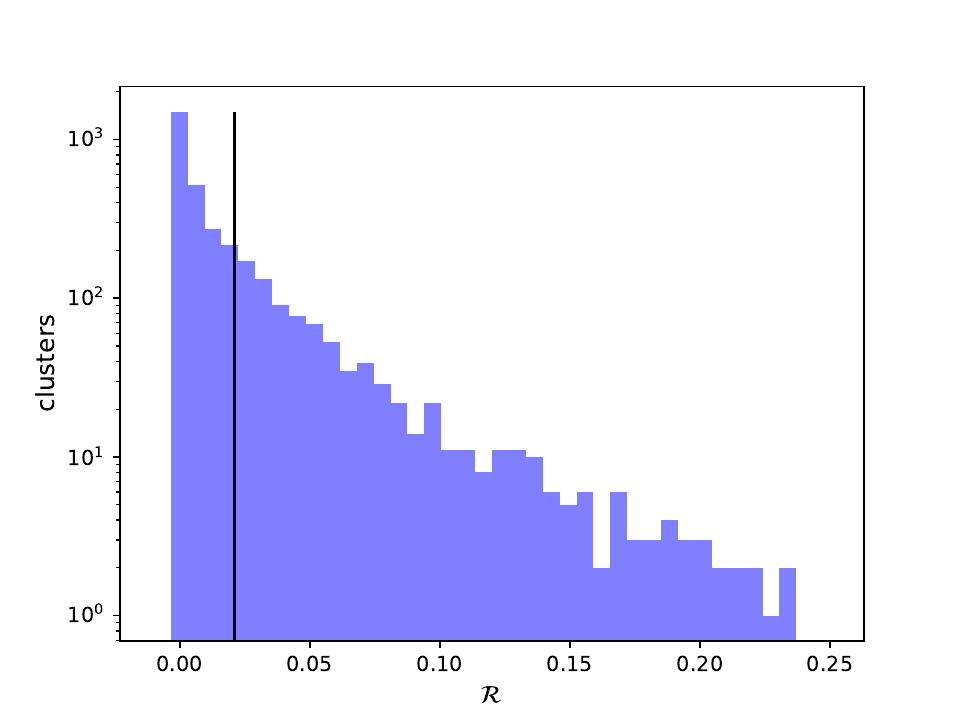}
        \caption{Distribution of the $\mathcal{R}$ values computed for the 3367 galaxy clusters in the merged catalogue outside quality~2 regions. The vertical line represents the value $\mathcal{R}^*_t=0.021$.}
        \label{fig:histR}
    \end{figure}

\begin{itemize}
    \item Name of the cluster;
    \item Coordinates in equatorial (J2000) coordinate systems;
    \item Redshift;
    \item $M_{500}$ and $R_{500}$;
    \item Alternative name derived from the cross-match between the four original catalogues;
    \item Name of the mosaic at the position of the cluster in the DR3 mosaicking scheme;
    \item The names of all pointings used to create the mosaic at the position of the cluster (i.e. names of mosaics in the DR2 mosaicking scheme);
    \item The distance of the cluster from the pointing centre of each pointing;
    \item Noise at the cluster location in each pointing before mosaicking;
    \item Quality class of each mosaic in the DR2 mosaicking scheme;
    \item Comments on the pointing used for the computation of $\mathcal{R}$;
    \item Name of the mosaic used to compute $\mathcal{R}$. For quality class~2 pointings, this is not listed since the $\mathcal{R}$ value is not computed;
    \item The value of $\mathcal{R}$ and its uncertainty, given on the basis of the cluster redshift as explained in Appendix~\ref{app:a};
    \item List of galaxy clusters which overlap with the system within their 2.2$R_{500}$.
\end{itemize}

The table is available on a dedicated page of the survey website\footnote{\url{https://lofar-surveys.org/clusters_dr3.html}}. The information provided in this table can be used to build a sub-sample of galaxy clusters for further studies (an example is provided in the following section).

It is also useful to derive global statistics for this large sample of galaxy clusters. The distribution of $\matchcal{R}$ values across the merged sample --restricted to the 3367 clusters located in regions where the segmentation maps are classified as quality~0 or~1 -- is shown in Fig.~\ref{fig:histR}. Using a threshold of $\mathcal{R} > 0.021$, we identify 918 galaxy clusters as candidates hosting diffuse radio emission, or DE. In this case, we do not exclude clusters based on the uncertainty of their $\mathcal{R}$ value, in order to retain the largest possible sample size, at the expense of a lower classification accuracy.

\begin{table}
    \centering
    \small
    \caption{Occurrence of galaxy clusters and detection fraction of the four galaxy cluster catalogues.}
    \begin{tabular}{ccccccc}
    \hline
    \hline
        Catalogue & $N_{\rm LoTSS-DR3}$ & $N_0$ & $N_1$ & $N_2$ & $N_{\mathcal{R}>0.021}$ & $\mathit{f}_{DE}$ \\
        \hline
         PSZ2 & 686 & 459 & 167 & 60 & 344 & $50\%$ \\
         ACT-DR5 & 1699 & 696 & 684 & 319 & 327 & $22\%$ \\
         MCXC2 & 874 & 611 & 188 & 75 & 283 & $32\%$ \\
         1eRASS & 1324 & 867 & 405 & 52 & 286 & $21\%$ \\
         \hline
    \end{tabular}
    \tablefoot{Column 1: Catalogue name. Column 2: Number of galaxy clusters within the LoTSS-DR3 area with $M_{500}>10^{14}M_{\odot}$. Column 3: Number of galaxy clusters observed in quality class~0 mosaics. Column 4: Number of galaxy clusters observed in quality class~1 mosaics. Column 5: Number of galaxy clusters observed in quality class~2 mosaics and therefore excluded from the analysis. Column 6: Number of galaxy clusters with $\mathcal{R}$ value larger than 0.021. Column 7: Intrinsic fraction of clusters with diffuse radio emission over the total number of clusters analysed for each catalogue, computed following Eq.~\ref{eq:occur}.}
    \label{tab:occurrence}
\end{table}

   \begin{figure}[]
        \centering
        \includegraphics[width=0.9\linewidth]{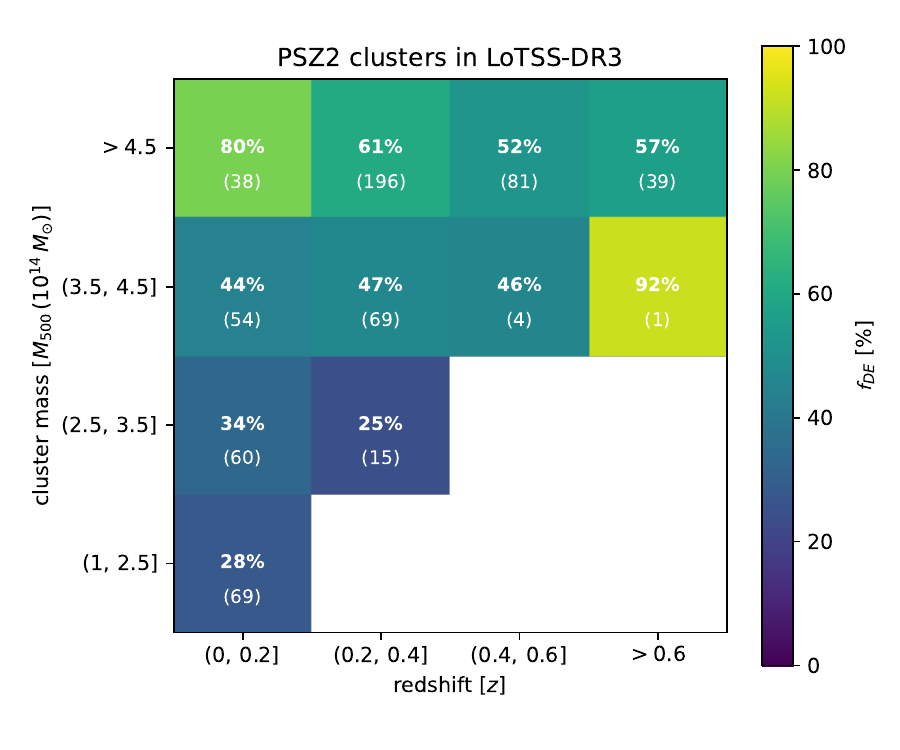}
        \caption{Intrinsic fraction of clusters with diffuse radio emission over the total number of clusters in each mass and redshift bin in the PSZ2 cluster catalogue computed following Eq.~\ref{eq:occur}. The total number of clusters in each bin ($N_0$+$N_1$) is also shown in brackets.}
        \label{fig:occurrence_PSZ2}
    \end{figure}

However, this number must be corrected for the classification performance of our detection pipeline. As discussed in Sec.~\ref{sec:detpipeline}, the threshold $\mathcal{R}_t^*=0.021$ corresponds to a precision of $68\%$ and a recall of $74\%$ when tested on the LoTSS-DR2/PSZ2 sample of 246 galaxy clusters. Under the assumption that these performance metrics are applicable to the LoTSS-DR3 data, we estimated the intrinsic fraction of galaxy clusters hosting diffuse radio emission, $\mathit{f}_{DE}$, as

\begin{equation}
    \rm \mathit{f}_{DE} = \frac{Precision(\mathcal{R}^*_t=0.021)}{Recall(\mathcal{R}^*_t=0.021)} \frac{N_{\mathcal{R}>0.021}}{N_0+N_1} \simeq 0.92 \frac{N_{\mathcal{R}>0.021}}{N_0+N_1} ~,
    \label{eq:occur}
\end{equation}

\noindent which yields $\mathit{f}_{\rm DE} \simeq 25\%$ for the merged catalogue. This estimate should be treated with caution, as both the adopted threshold and the corresponding precision and recall are derived from a test sample based on LoTSS-DR2/PSZ2 galaxy clusters. Given that the associated uncertainties are dominated by systematic effects related to sample selection and transferability of the classifier performance, we do not attempt to provide a formal error estimate for $\mathit{f}_{\rm DE}$.

In order to derive statistical information on the detection rate within each catalogue, it is convenient to associate each cluster with its original galaxy cluster catalogue. This association is performed using the alternative names listed in the merged catalogue. Table~\ref{tab:occurrence} reports the detection rate for each of the four galaxy cluster catalogues. The total number of clusters exceeds that of the merged catalogue, as several clusters appear in more than one catalogue. 

The intrinsic fraction of galaxy clusters hosting diffuse radio emission is highest in the PSZ2 catalogue ($50\%$). This is expected, as this catalogue preferentially includes more massive and lower-redshift systems compared to the other three catalogues, as shown in Fig.~\ref{fig:clusters_all}. Both theoretical models for the origin of radio halos \citep{Cassano07} and cosmological simulations investigating the formation of radio relics \citep{Lee24} predict that the fraction of galaxy clusters hosting diffuse radio sources increases with increasing cluster mass and decreasing redshift, and this is confirmed by observational results in the case of radio halos \citep{Cassano23}.

We test this expectation further by dividing each cluster catalogue into mass and redshift bins and computing the intrinsic fraction of galaxy clusters hosting diffuse radio emission in each bin. The results for the PSZ2 catalogue are shown in Fig.~\ref{fig:occurrence_PSZ2}, while the corresponding plots for the other three catalogues are presented in Fig.~\ref{fig:occurrence_others}. Apart from statistical fluctuations in bins containing a small number of clusters (with the total number of clusters in each bin shown in brackets), we recover a clear trend of increasing detection fraction towards higher masses and lower redshifts for all catalogues.

\citet{Botteon22} found that, in the LoTSS-DR2 sample, the detection rate of galaxy clusters hosting a radio halo is $30\pm11\%$, while it is $10\pm6\%$ for clusters hosting one or more radio relics. They also reported that $51\pm14\%$ of the clusters do not show evidence of diffuse synchrotron emission in the ICM. This is closely consistent with the intrinsic detection fraction of $50\%$ that we have obtained for the PSZ2 sample, where radio halos, relics and other diffuse radio sources of uncertain classification are accounted for. Furthermore, they predicted the detection of $251\pm92$ galaxy clusters hosting a radio halo and $83\pm50$ clusters hosting at least one relic from the PSZ2 catalogue in the full LoTSS survey. Following our approach we have found 344 galaxy clusters hosting diffuse radio emission in the LoTSS-DR3/PSZ2 sample. Although a direct comparison among these numbers should be treated with caution, the overall fraction of detected systems is broadly consistent with these predictions.

\subsection{Curated sub-sample of galaxy clusters detected by the network}
\label{sec:goldsample}

Finally, we have used the merged catalogue to select a sub-sample of galaxy clusters having the highest classification accuracy provided by our method. We have selected all clusters in the merged catalogue with (i) quality class~0 (ii) $\mathcal{R}>0.021$, and (iii) a reliable classification flag with respect to this threshold, meaning that $\mathcal{R} - 0.021 > \sigma$, where $\sigma$ is the uncertainty on the $R$ value discussed in App.~\ref{app:a}. This set contains 357 galaxy clusters. The table and the images of this sub-sample are available on a dedicated page of the survey website\footnote{\url{https://lofar-surveys.org/clusters_dr3.html}}

For this sub-sample, we performed a visual inspection of the LoTSS-DR3 images at both high and low angular resolution, aimed at identifying clear false detections through a comparison with the segmentation maps produced by Radio U-Net. Clear false detections were identified when the regions segmented by the network overlapped with extended AGN emission, nearby galaxies, or other foreground or background galaxy clusters projected along the line of sight. We found a total of 19 clear false detections. 

We subsequently searched the literature to identify previous studies of each cluster. In total, we identified 109 galaxy clusters with literature confirmation of the presence of diffuse radio emission, in the form of radio halos, mini-halos, radio relics, phoenixes, remnant radio sources, or diffuse radio sources with uncertain classification. 

Out of the 357 clusters, 104 clusters belong to the PSZ2 catalogue and lie within the LoTSS-DR2 area already studied by \citet{Botteon22}. However, 31 of these systems were not classified as hosts of diffuse radio emission in this previous work, either because of the low quality of the available data (9 clusters) or because the detected emission was considered by \citet{Botteon22} to be unrelated to the ICM (22 clusters, but in three of these cases the emission was later reclassified as a remnant radio source by \citealt{Bruno25}). Seven clusters belong instead to the sample of non-PSZ2 systems found to host diffuse radio emission by \citet{Hoang22}. Interestingly, both inter-cluster radio bridges discovered with LOFAR -- the Abell~399–401 system \citep{Govoni19} and Abell~1758 \citep{Botteon20c} -- are included in our curated sub-sample.

Among the remaining 229 galaxy clusters lacking previous studies in the literature, several show radio morphologies that can be readily associated with standard classes of diffuse radio emission, such as radio halos or radio relics. Examples of these cases are shown in Fig.~\ref{fig:example}. Follow-up studies are already ongoing on a few of these systems, such as the high-redshift galaxy cluster \object{PSZ2 G209.79+10.23} ($z$=0.677), possibly hosting a radio halo and double radio relics (Di Gennaro et al., in prep.), or the possible double radio relic system in \object{ACT-CL J0245.1+1843} (Pfeifer et al., in prep.). Another peculiar case is represented by three galaxy clusters in the curated sub-sample that belong to a single interacting system around \object{PSZ2 G343.33+83.19}, in which the network detected diffuse radio emission connecting the three clusters (see Fig.~\ref{fig:peculiar}). For most of the remaining candidate new diffuse radio sources, dedicated source extraction and follow-up analyses of individual objects are required to confirm the association of the diffuse radio emission with the ICM and to determine its physical nature.

In summary, the curated sub-sample comprises 357 galaxy clusters, of which 19 are clear false detections and 109 have been previously confirmed to host some form of diffuse radio emission in the literature. The remaining 229 clusters will be the subject of dedicated post-processing and follow-up investigations.

We note that this sub-sample does not represent a complete census of all galaxy clusters with diffuse radio emission identified by the network, but it was selected in order to ensure the highest accuracy achievable by the network ($76\%$), according to the performance metrics discussed in Sec.~\ref{sec:detpipeline}. Some clusters may have been excluded either because they are located in regions classified as quality class~1 or~2, or because their $\mathcal{R}$ value lies close to the adopted threshold. This is, for example, the case for \object{PSZ2~G135.76–62.03} (Abell~0168; Hoang et al., in prep.), \object{PSZ2~G075.71+13.51} (Abell~2319; Drabent et al., in prep.), and \object{PSZ2~G200.82+27.42} and \object{ACT-CL~J0220.9+0652} (Pfeifer et al., in prep.), which are detected with high significance but fall within quality class~1 regions, and for which dedicated follow-up studies are already in progress. 


\section{Discussion}
\label{sec:discussion}

In this section, we discuss the results obtained in this paper. In particular, we focus on the use of the catalogues produced in this work and on possible future improvements of the pipeline.

\subsection{Towards statistical studies of galaxy clusters}

This work is meant to provide automatic diffuse radio source detection in the largest possible sample of galaxy clusters and to guide blind detection of new diffuse radio sources. These kinds of automated methodologies are becoming increasingly important with new-generation wide-area surveys, which will allow statistical studies based on hundreds of sources. With our method, we obtained a trend of an increasing detection rate of diffuse radio sources with increasing cluster mass and decreasing redshift (Sec.~\ref{sec:results1}), which is particularly encouraging, as it reproduces behaviours observed in statistical samples \citep{Cassano23}. However, we recognise that the creation of a statistically meaningful sub-sample of galaxy clusters  \citep[such as those based on the LoTSS-DR2/PSZ2 sample; e.g.][]{Bruno23a,Cassano23,Cuciti23,Jones23},  requires additional efforts. This includes, for example, cleaning the present catalogues from spurious detections, comparing detections with multi-wavelength datasets, and carefully measuring the flux of diffuse radio emission.

In order to create a subsample of galaxy clusters for scientific purposes, a decision should be made on the $\mathcal{R}$ value threshold to be used as a parameter for detection: higher $\mathcal{R}$ values will lead to purer samples, with a smaller fraction of DE objects but detected with higher significance, while lower $\mathcal{R}$ values will lead to more complete sample but with a higher fraction of false detection. In Sec.~\ref{sec:detpipeline} we selected a sub-sample with the highest possible accuracy which is $76\%$. In this curated sub-sample, the precision is $74\%$, as derived from the test carried out on the LoTSS-DR2/PSZ2 sample, meaning that the $74\%$ of clusters classified as DE by the network are true detections. While the implementation of more sophisticated automated pipelines is desirable, this is the best achievable compromise between automation and accuracy at present.

After this selection, in many cases, a detection can still be confirmed only by the comparison with multi-wavelength datasets, for example, to discriminate between diffuse radio emission connected to AGN and related to the ICM, for which the comparison with optical datasets is necessary. This is even more important for classification purposes, for example, regarding the position of the diffuse radio emission with respect to the X-ray emission of the cluster to discriminate between relics and halos. Additionally, radio spectral index information should be used to discriminate between different acceleration mechanisms at work in different sources. This will be available in a wide area thanks, for example, to LOFAR observation in the low (54 MHz) and high (144 MHz) bands. While the classification of sources could remain uncertain even after these comparisons, these are clearly important steps to be taken for subsequent studies.

Furthermore, as mentioned in Sec.~\ref{sec:lotss}, the automated calibration and imaging strategy of LoTSS is not optimised for the diffuse radio emission of galaxy clusters. For the purpose of this work -- which is also minimising re-procesing -- we have directly used LoTSS images at $20\arcsec$ as input for the pipeline, where diffuse radio emission could be incompletely deconvolved or affected by the presence of residual calibration and imaging artefacts. When a sub-sample is selected, dedicated pipelines should be used for source extraction and self-calibration of each target \citep{vanWeeren21, Botteon22}, which allows a more robust characterisation of their morphology and measurement of their fluxes. However, as already noted in \citet{Stuardi24} and further explored in \citet{Stuardi25}, in some cases the network is able to detect diffuse radio emission that is not detected in extracted images, probably because they are at the detection limit, but emerges in deeper observations or lower-resolution images. Hence, putting together machine learning information and tailored post-processing of selected candidates is still essential.

\subsection{Small angular size systems}

The accuracy metrics reported for the current catalogues are all based on the LoTSS-DR2/PSZ2 sample, which contains more low-redshift and high-mass systems with respect to other galaxy cluster catalogues (see Appendix~\ref{app:allclusters}, Fig.~\ref{fig:singlecat}). As explored in Appendix~\ref{app:a}, the variability of the $\mathcal{R}$ value increases for small angular size systems, i.e. high-redshift and low-mass clusters, and consequently, the accuracy metrics could vary in different samples. Hence,  the computed $R$ value for high-redshift and low-mass systems should be taken with caution.

In particular, the detection of diffuse radio sources at high redshifts is more difficult because of redshift dimming and increasing inverse Compton energy losses, which make the emission fainter and harder to separate from contaminating sources. Moreover, the majority of high-redshift clusters have fairly low masses ($M_{500}\lesssim5\times10^{14}~{\rm M_\odot}$), where the presence of diffuse radio emission is less likely \citep{DiGennaro21a,Cassano23,DiGennaro25}. In many cases, only increasing the angular resolution of LOFAR observations would help firmly establish the presence of the diffuse radio emission detected by Radio U-Net. Increasing the angular resolution to the sub-/arcsecond regime is now possible with LOFAR's international baselines. This has been helpful in confirming the presence of diffuse radio emission in clusters \citep{Hlavacek25,DiGennaro26}. In the future, dedicated studies of a large sample of high-redshift clusters may help better train the network in the high-redshift regime.

\subsection{Machine learning methodology improvements}
\label{sec:discuss_ML}

Strategies to enhance the performance of DL methods applied to the detection of diffuse and extended radio sources are currently being explored by several of the authors. A first approach focuses on improving the network architecture itself, in particular by incorporating attention mechanisms. In this context, attention layers allow the network to dynamically weight different regions of the input image, enabling it to focus on spatial features that are more relevant for the task at hand while suppressing less informative or contaminating structures. The effectiveness of this approach has already been demonstrated by the TUNA network \citep{Sanvitale25}. Additional improvements may be achieved by fine-tuning the models on real observational data, a procedure already explored in \cite{Stuardi24} and currently being further tested on TUNA. 

A second strategy targets improvements to the training data, aiming at producing more realistic and comprehensive mock observations. This includes incorporating radio galaxies either by extracting them directly from real observations or by using new generations of cosmological simulations in which AGN activity and star formation are modelled self-consistently. Furthermore, training data can be enhanced by explicitly accounting for Galactic foreground emission and other large-scale diffuse components, which may otherwise be misclassified as cluster-related emission. This would improve the generalisation capabilities of the network.

An additional research direction under investigation involves a multi-wavelength approach, in which radio, X-ray, and optical observations are jointly used as network inputs. This strategy has already proven effective in the context of galaxy cluster identification and classification in X-ray surveys, where machine learning techniques have been successfully employed to distinguish between relaxed and disturbed systems or to identify galaxy clusters \citep{Kosiba20, Sadikov25}. Extending this approach to radio studies may help the network better isolate emission physically associated with the ICM.

Beyond supervised learning, self-supervised and semi-supervised methods are also being explored as a way to mitigate the limited availability of labelled data. Recent studies \citep[e.g.][]{Mostert21,Mostert23,Slijepcevic24,BaronPerez25} have shown that such approaches can significantly improve performance and robustness, particularly for rare and complex radio sources. Self-supervised methodologies will be even more important in the SKA era, when new sources with previously unknown morphologies could be discovered.

\section{Conclusions}
\label{sec:conclusions}

We have presented the first catalogue of diffuse radio emission in galaxy clusters extracted from LoTSS-DR3 based on DL. By combining the unprecedented sky coverage of LoTSS with the Radio U-Net convolutional neural network, we have developed an automated pipeline capable of identifying extended low-surface-brightness emission related to the ICM.

Our analysis processed 2551 LoTSS-DR3 mosaics and cross-matched the resulting segmentation maps with 3822 clusters from four complementary catalogues (PSZ2, ACT-DR5, MCXC2, and 1eRASS). The method provides both a quantitative indicator for the presence of diffuse radio emission in each cluster ($\mathcal{R}$ value) and a pixel-level segmentation map of the entire survey. The detection fraction of diffuse radio sources in the four catalogues increases in high-mass and low-redshift systems, as predicted by current models and simulations. 

From the same catalogue, we also curated a sub-sample of galaxy clusters with a reliable classification flag containing 357 galaxy clusters showing significant evidence of diffuse radio emission. Dedicated pipelines should be used for source extraction and self-calibration of each new target in this sub-sample \citep{vanWeeren21, Botteon22}, to allow for a more robust characterisation of their morphology and measurement of their fluxes.

Although our approach does not yet distinguish between different types of diffuse radio emission (e.g. radio halos, relics, or phoenixes and remnant plasma sources), it establishes a framework for automatic research of diffuse radio emission sources in wide-area radio surveys. The same procedure can also be applied to different radio surveys by tailoring the training dataset to the specific telescope.

Limitations regarding our method remain, primarily related to the network’s ability to generalise in the presence of Galactic diffuse radio emission, imaging artefacts, and diffuse radio emission associated with extended AGN, which can lead to false detections. In future developments, we will focus on retraining the Radio U-Net using more realistic simulations and an expanded set of real observational data.

The catalogue released with this paper provides a valuable resource for the community, enabling follow-up studies of individual systems, statistical analyses, and cross-correlation with multi-wavelength datasets. The upcoming SKA and LOFAR2.0 surveys will dramatically expand the available data volume, and methodologies such as the one presented in this work will be essential for exploiting their full scientific potential.

\section*{Data availability}

The processed data products and catalogues generated in this work are publicly available on a dedicated page of the LOFAR Surveys website at \url{https://lofar-surveys.org/clusters_dr3.html}. These include three FITS tables with information on: (i) the full sample of 3822 galaxy clusters in the merged catalogue, (ii) the 357 clusters in the curated sub-sample, and (iii) all LoTSS-DR3 mosaics. A HiPS file showing the segmentation map for the entire LoTSS-DR3 (excluding quality~2 mosaics) is also available at the same location, while segmentation maps of individual mosaics are provided upon reasonable request. The catalogues are also
available at the CDS and on Zenodo \url{https://zenodo.org/records/20022649}.

The LoTSS-DR3 data underlying this article are publicly available from the LOFAR Surveys webpage at \url{(https://lofar-surveys.org/dr3_release.html}). The Radio U-Net code used in this work is available at \url{https://github.com/ICSC-Spoke3/Radio-U-Net}. The simulations used to train the network are available at \url{https://owncloud.ia2.inaf.it/index.php/s/IbFPlCCcPUresrr} and are described by \citet{Stuardi24}.

\begin{acknowledgements}
This paper is supported by the Fondazione ICSC, Spoke 3 Astrophysics and Cosmos Observations -- National Recovery and Resilience Plan (Piano Nazionale di Ripresa e Resilienza, PNRR) Project ID CN\_00000013 "Italian Research Center for High-Performance Computing, Big Data and Quantum Computing" funded by MUR Missione 4 Componente 2 Investimento 1.4: Potenziamento strutture di ricerca e creazione di "campioni nazionali di R\&S (M4C2-19)" - Next Generation EU (NGEU), and it's also supported by (Programma Operativo Nazionale, PON), ``Tematiche di Ricerca Green e dell'Innovazione". The results of this work have been produced on the Leonardo Supercomputer at CINECA (Bologna, Italy) in the framework of the ISCRA project IsCc2 COS4ML (P.I. C.Stuardi) and also supported by the INAF Mini Grant (P.I C.Stuardi) awarded by the "Bando Ricerca Astrofisica Fondamentale 2024". We also acknowledge the usage of online storage tools kindly provided by the INAF Astronomical Archive (IA2) initiative (\url{http://www.ia2.inaf.it}). \\
LOFAR \citep{vanHaarlem13} is the Low Frequency Array designed and constructed by ASTRON. It has observing, data processing, and data storage facilities in several countries, which are owned by various parties (each with their own funding sources), and which are collectively operated by the LOFAR ERIC under a joint scientific policy. The LOFAR resources have benefited from the following recent major funding sources: CNRS-INSU, Observatoire de Paris and Université d'Orléans, France; BMFTR, MKW-NRW, MPG, Germany; Science Foundation Ireland (SFI), Department of Business, Enterprise and Innovation (DBEI), Ireland; NWO, The Netherlands; The Science and Technology Facilities Council, UK; Ministry of Science and Higher Education, Poland; The Istituto Nazionale di Astrofisica (INAF), Italy. This research made use of the Dutch national e-infrastructure with support of the SURF Cooperative (e-infra 180169) and the LOFAR e-infra group. The Jülich LOFAR Long Term Archive and the German LOFAR network are both coordinated and operated by the Jülich Supercomputing Centre (JSC), and computing resources on the supercomputer JUWELS at JSC were provided by the Gauss Centre for Supercomputing e.V. (grant CHTB00) through the John von Neumann Institute for Computing (NIC). This research made use of the University of Hertfordshire high-performance computing facility and the LOFAR-UK computing facility located at the University of Hertfordshire and supported by STFC [ST/P000096/1], and of the Italian LOFAR-IT computing infrastructure supported and operated by INAF, including the resources within the PLEIADI special ``LOFAR'' project by USC-C of INAF, and by the Physics Department of Turin university (under an agreement with Consorzio Interuniversitario per la Fisica Spaziale) at the C3S Supercomputing Centre, Italy. This research is part of the project LOFAR Data Valorization (LDV) [project numbers 2020.031, 2022.033, and 2024.047] of the research programme Computing Time on National Computer Facilities using SPIDER that is (co-)funded by the Dutch Research Council (NWO), hosted by SURF through the call for proposals of Computing Time on National Computer Facilities. A.Bon. and M.B. acknowledge support from the ERC CoG $\vec{B}$ELOVED, N.101169773. F.dG., G.D.G., N.B. and M.C. acknowledge support from the ERC CoG ULU, N. 101086378. M.J.H. thanks the UK STFC for support [ST/V000624/1, ST/Y001249/1]. F.G. acknowledges the financial contribution from the INAF GO grant 1.05.24.02.10 Extended Radio Emission in Galaxy Clusters at deep focus with MeerKAT. F.V. acknowledges funding under the European Union’s Horizon Europe program through the ERC Synergy Grant COSMOMAG (Project Id. 101224803).

\end{acknowledgements}

\bibliographystyle{aa}
\bibliography{my_bib}

\begin{appendix}

\section{Single cluster catalogues and detection fraction}
\label{app:allclusters}

\begin{figure}
\centering
\includegraphics[width=0.42\textwidth]{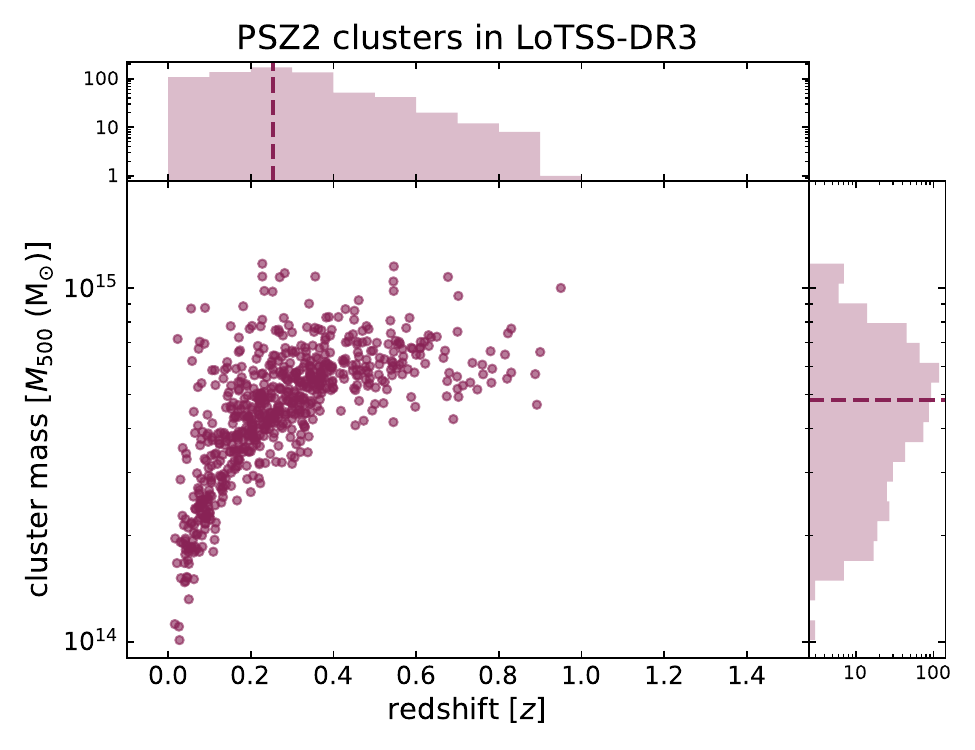}
\includegraphics[width=0.42\textwidth]{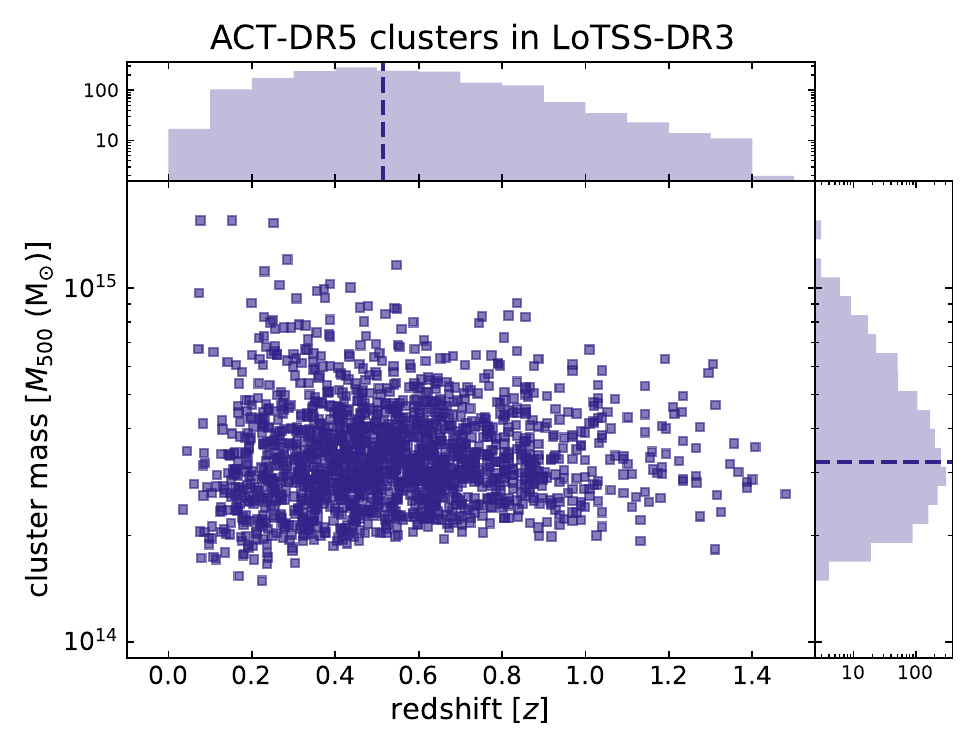}
\includegraphics[width=0.42\textwidth]{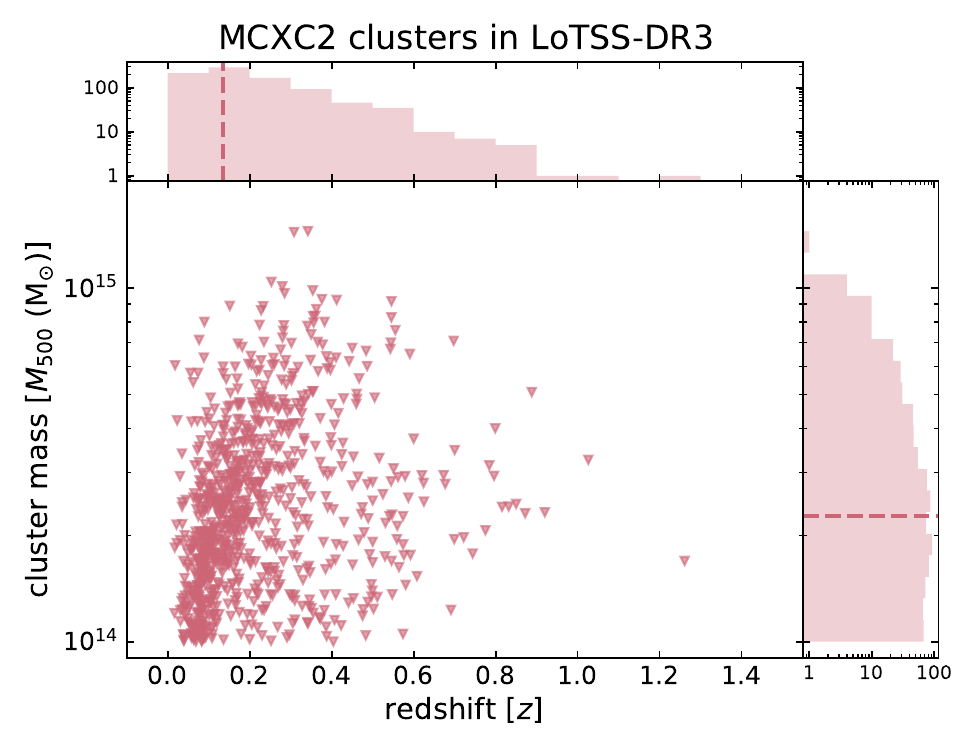}
\includegraphics[width=0.42\textwidth]{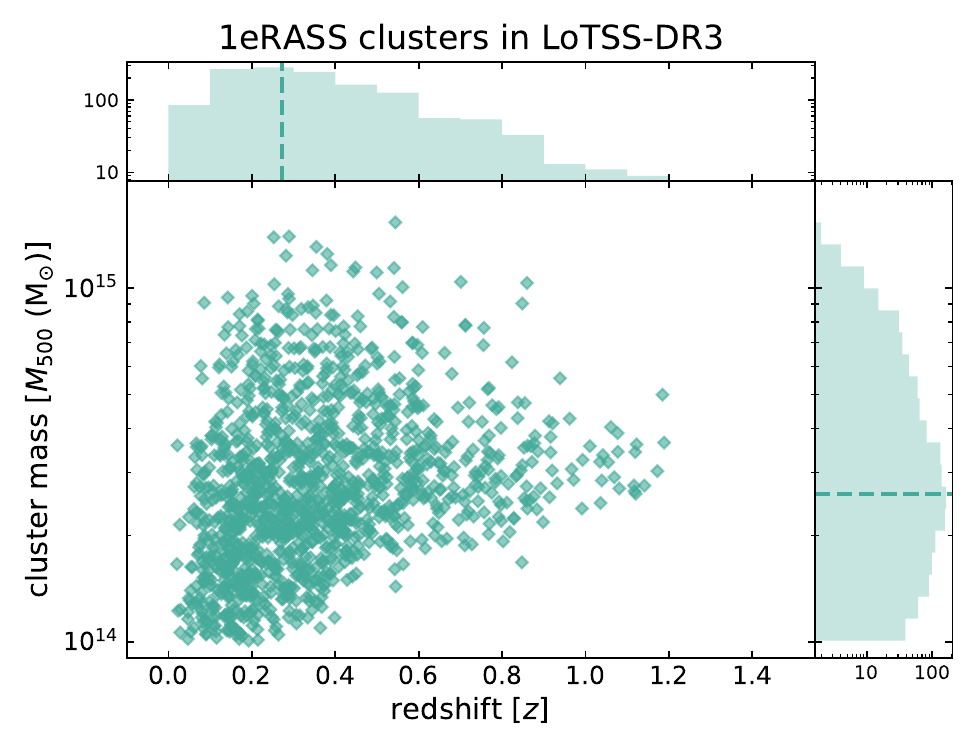}    
\vspace{-5mm}
\caption{Mass versus redshift plots of the single selected cluster catalogues in LoTSS-DR3. Dashed lines in the horizontal and vertical histograms show the median values of the redshift and mass, respectively.}
\label{fig:singlecat}
\end{figure}

\begin{figure*}
        \centering
        \includegraphics[width=\linewidth]{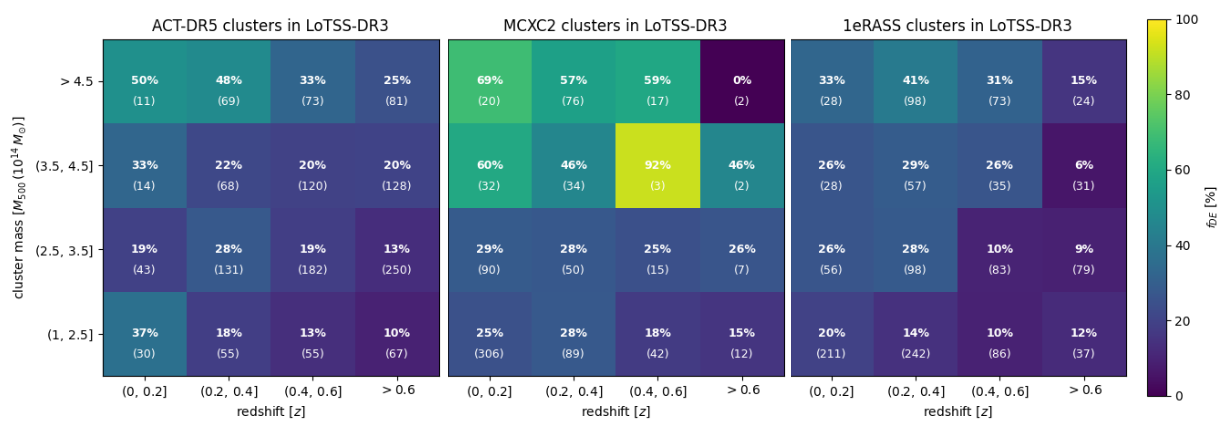}

        \caption{Intrinsic fraction of clusters with diffuse radio emission in the ACT-DR5, MCXC2 and 1eRASS catalogues computed following Eq.~\ref{eq:occur} and plotted for different mass and redshift bins. The total number of clusters in each bin (in quality~0 and quality~1 mosaics) is also shown in brackets.}
        \label{fig:occurrence_others}
\end{figure*}

In this Appendix, we show the mass-redshift diagram of the single cluster catalogues used to generate the merged catalogue in Sec.~\ref{sec:catalog} (Fig.~\ref{fig:singlecat}). We also show the intrinsic fraction of clusters detected with diffuse radio emission in the ACT-DR5, MCXC2 and 1eRASS catalogues, separated in mass and redshift bins as explained in Sec.~\ref{sec:results1} (Fig.~\ref{fig:occurrence_others}).

\section{Uncertainty on the $\mathcal{R}$ value}
\label{app:a}

In this work, we slightly modified the detection pipeline compared to \citet{Stuardi24} by applying Radio U-Net to the entire LoTSS-DR3 mosaics, rather than to individual cutouts centred on each cluster. The $\mathcal{R}$ values computed for the 246 LoTSS-DR2/PSZ2 galaxy clusters used as a test sample by \citet{Stuardi24} can therefore serve as a benchmark for the updated pipeline. Comparing the $\mathcal{R}$ values obtained with the two methods also allowed us to verify the robustness of the $\mathcal{R}$ parameter and to derive an appropriate statistical uncertainty.

\begin{figure}
        \centering
        \includegraphics[width=0.9\linewidth]{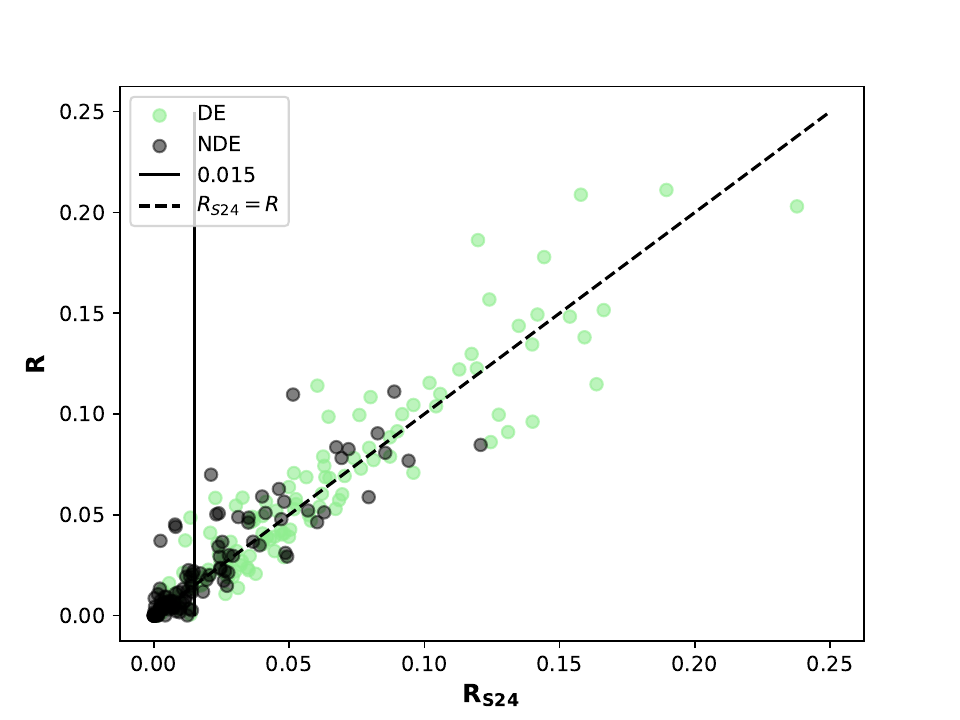}
        \caption{Comparison of the $\mathcal{R}$ value computed by \citet{Stuardi24} with the one computed in this work for the test sample of LoTSS-DR2/PSZ2 clusters. The vertical line marks the threshold used for classification by \citet{Stuardi24}.}
        \label{fig:R_dr2dr3}
    \end{figure}

In Fig.~\ref{fig:R_dr2dr3}, we plot the $\mathcal{R}$ values calculated in this work against those from \citet{Stuardi24}, denoted as $\mathcal{R}_{S24}$, for the 246 clusters classified as DE or NDE by \citet{Botteon22}. As expected, the two measurements follow a clear linear trend, and most clusters remain in the same class (identified using the threshold $\mathcal{R}_{S24}=0.015$ here). However, in some cases, significant differences in $\mathcal{R}$ are evident.

\begin{figure}
        \centering
        \includegraphics[width=0.9\linewidth]{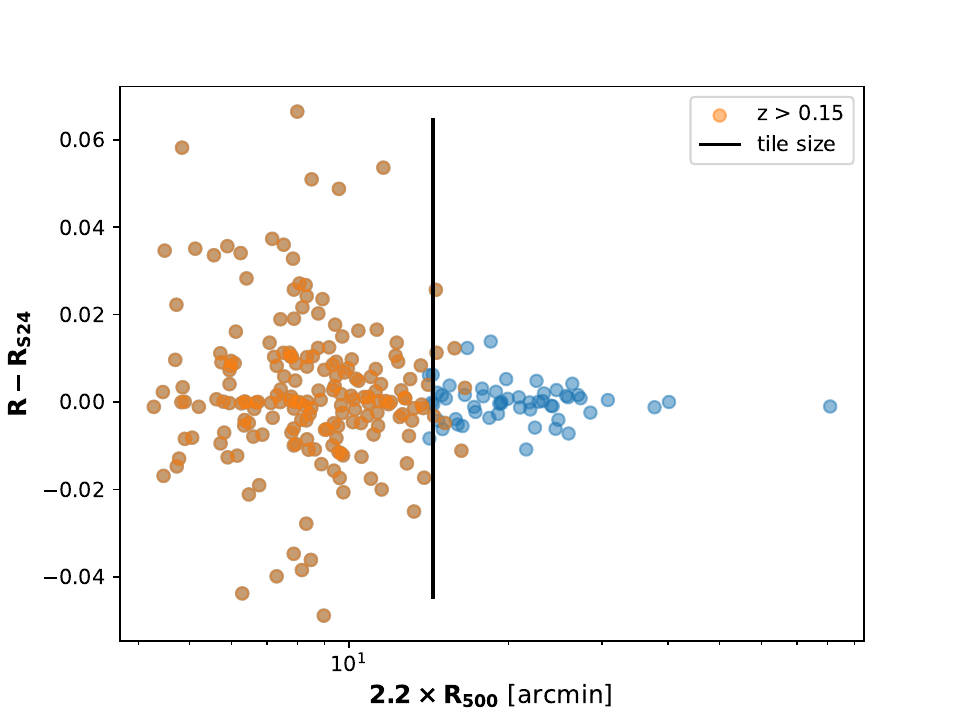}
        \caption{Difference between the $\mathcal{R}$ value computed in this work and in that of \citet{Stuardi24} for the test sample of LoTSS-DR2/PSZ2 clusters as a function of the radius used to compute $\mathcal{R}$. The size of the tile used by Radio U-Net is shown by the vertical line.}
        \label{fig:diffR_R}
    \end{figure}

We find that these discrepancies are statistically more pronounced for cluster with small projected angular size (see Fig.~\ref{fig:diffR_R}). The main cause of the discrepancy is the inference uncertainty associated with the model. Indeed, Radio U-Net is used here as a deterministic model. However, when it is applied for inference on images with different tiling schemes and varying noise properties, it will produce variability in the predictions for the same sources. This effect is expected to be more severe at high redshift where the diffuse radio emission is intrinsically fainter and difficult to detect. Small changes in the input can lead to large differences in the network’s prediction and, since the $\mathcal{R}$ parameter is, by definition, derived directly from the extent of the diffuse radio emission, and this emission at higher redshift is expected to span only a small number of pixels, a few FP or FN detections or shifts in the predicted values can produce significant differences in $\mathcal{R}$. 

    \begin{figure}
        \centering
        \includegraphics[width=0.9\linewidth]{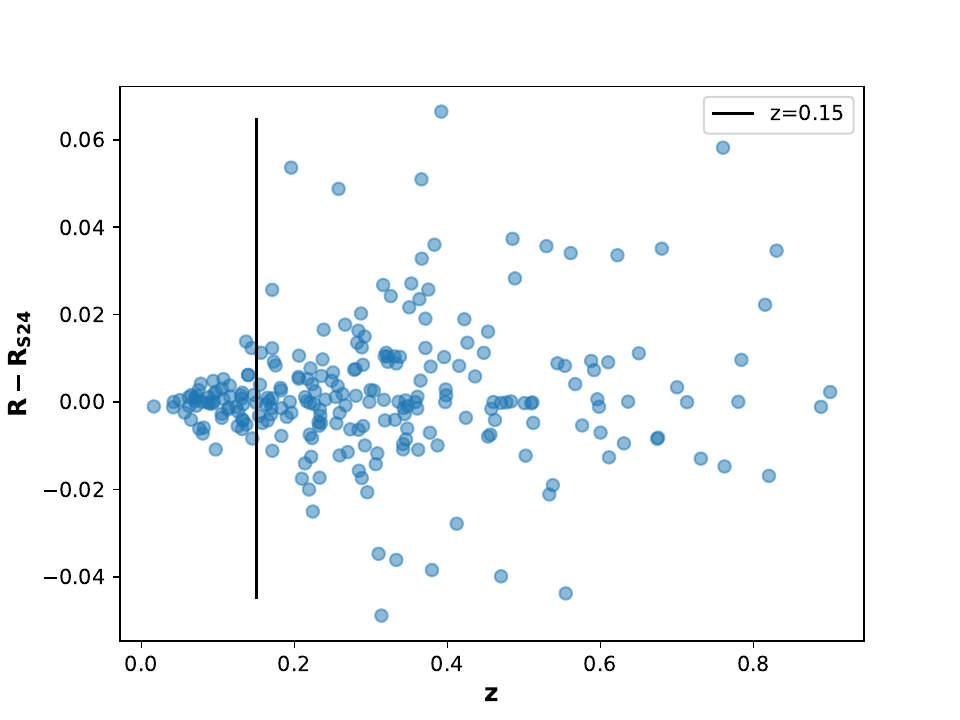}
        \caption{Difference between the $\mathcal{R}$ value computed in this work and in that of \citet{Stuardi24} for the test sample of LoTSS-DR2/PSZ2 clusters as a function of the clusters redshift.}
        \label{fig:diffR_z}
    \end{figure}

Another aspect to consider is the absence of high-redshift sources in the simulation set used to train the network, where diffuse radio emission was projected into mock lightcones up to $z=0.15$ \citep{Vazza19}. This could reduce the inference performances of the network for $z>0.15$ (see also Fig.~\ref{fig:diffR_z}). Another contributing factor is the choice of tile size for the segmentation: the tiles are designed to contain only a small fraction of the diffuse radio emission at $z<0.15$. When the radius used to compute $\mathcal{R}$ is smaller than the tile size, the $\mathcal{R}$ value becomes sensitive to the relative position of the tile and the diffuse radio emission. \citet{Stuardi24} performed segmentation on images centred on the cluster positions, whereas in this work, the segmentation was carried out on entire pointings, meaning the tiles were randomly centred with respect to the clusters. This effect is illustrated in Fig.~\ref{fig:diffR_R}, which also shows that cluster size and redshift are correlated, suggesting that both the under-representation of high-$z$ systems in the training set and the choice of tile size could contribute to increased uncertainty in the $\mathcal{R}$ values for small angular size systems.

To account for this, we computed a statistical uncertainty on $\mathcal{R}$ as a function of cluster redshift. The test sample was divided into five redshift bins, each containing 49 clusters, and for each bin we calculated the standard deviation of the absolute differences between the two $\mathcal{R}$ estimates:
\begin{equation}
    \sigma_{z_{\rm bin}} = \sigma\!\left(|\mathcal{R}-\mathcal{R}_{S24}|_{z_{\rm bin}}\right),
\end{equation}
where the five bins are listed in Tab.~\ref{tab:uncert} with corresponding uncertainties.

\begin{table}
    \centering
    \small
    \caption{Uncertainties on the $\mathcal{R}$ value computed in each redshift bin.}
    \begin{tabular}{cc}
    \hline
    \hline
        $z_{bin}$ & $\sigma_{z_{bin}}$ \\
        \hline
        $z \leq 0.1394$ & 0.004  \\
        $z \in (0.1394,0.23225]$ & 0.012 \\
        $z \in (0.23225,0.321677]$ & 0.015 \\
        $z \in (0.321677,0.4561]$ & 0.019  \\
        $z > 0.4561$ & 0.019 \\
         \hline
    \end{tabular}
    \label{tab:uncert}
\end{table}
 
These uncertainties are assigned to the corresponding $\mathcal{R}$ values in the cluster catalogues produced in this work.

\section{Examples of galaxy clusters from the curated sub-sample}
\label{app:examples}

\begin{figure}
    \centering
    \includegraphics[width=0.525\textwidth]{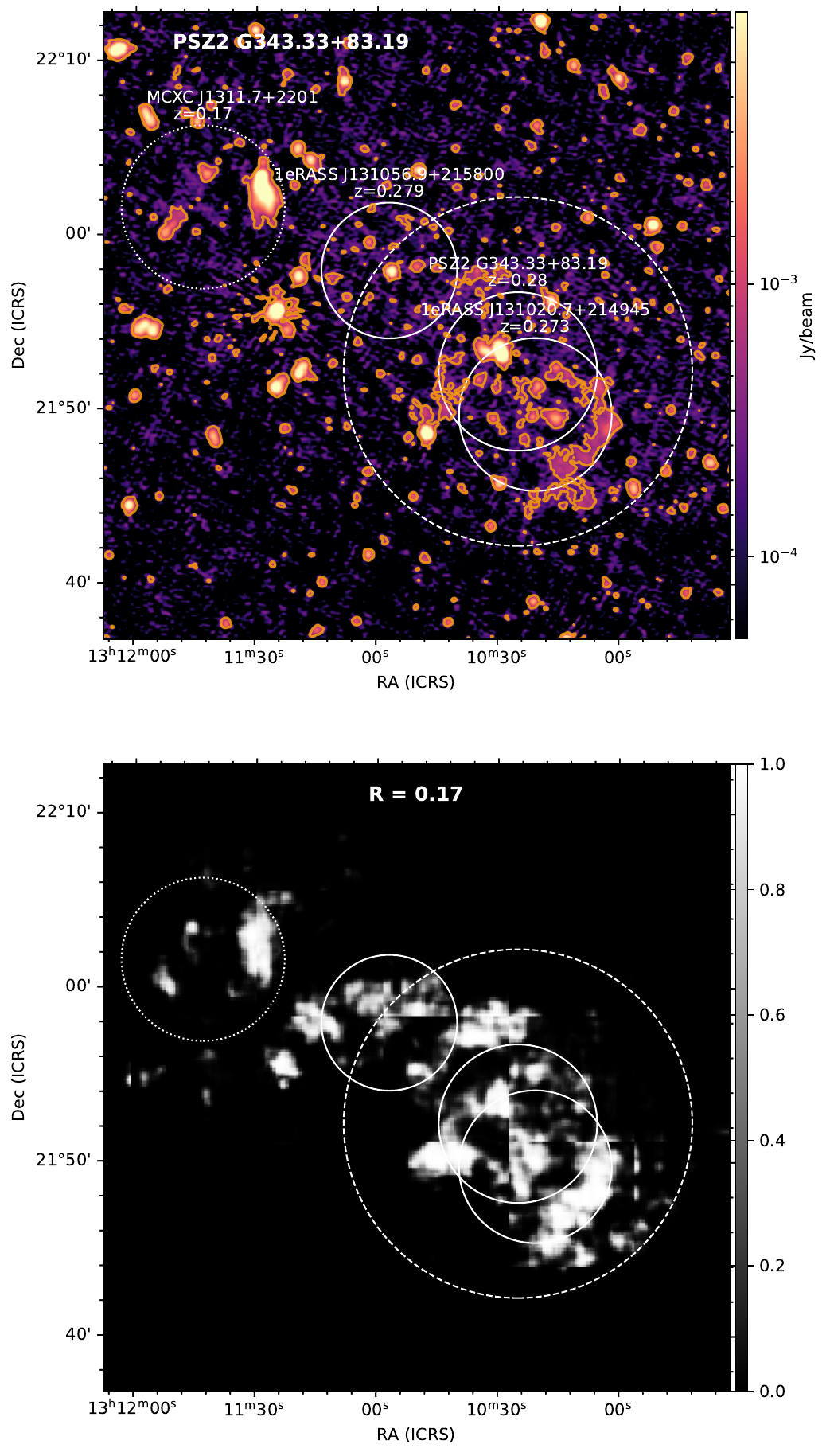}
    \caption{Example of a peculiar system within the curated sub-sample. The LoTSS-DR3 image at $20\arcsec$ resolution is shown in the top panel, with contours at the $3\sigma$ level (0.39 mJy/beam). The three solid white circles show the $R_{500}$ of the three clusters found at similar redshifts. The dashed circle shows the 2.2$R_{500}$ radius of the main cluster (\object{PSZ2 G343.33+83.19}), while the dotted circle shows another cluster in the foreground. The segmentation map is shown in the bottom panel.}
    \label{fig:peculiar}
\end{figure}

\begin{figure*}
        \centering
        \includegraphics[width=0.9\linewidth]{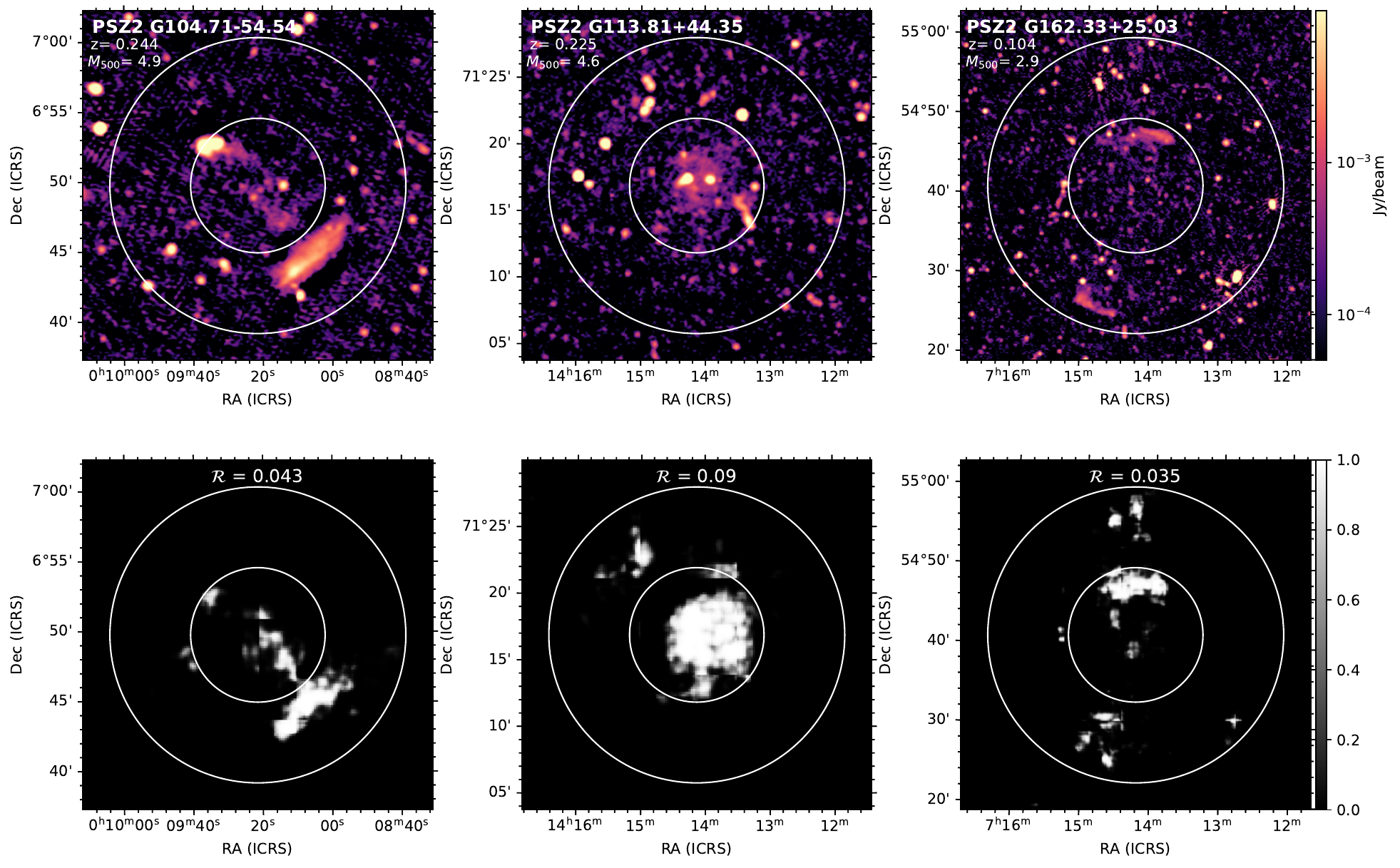}
        \includegraphics[width=0.9\linewidth]{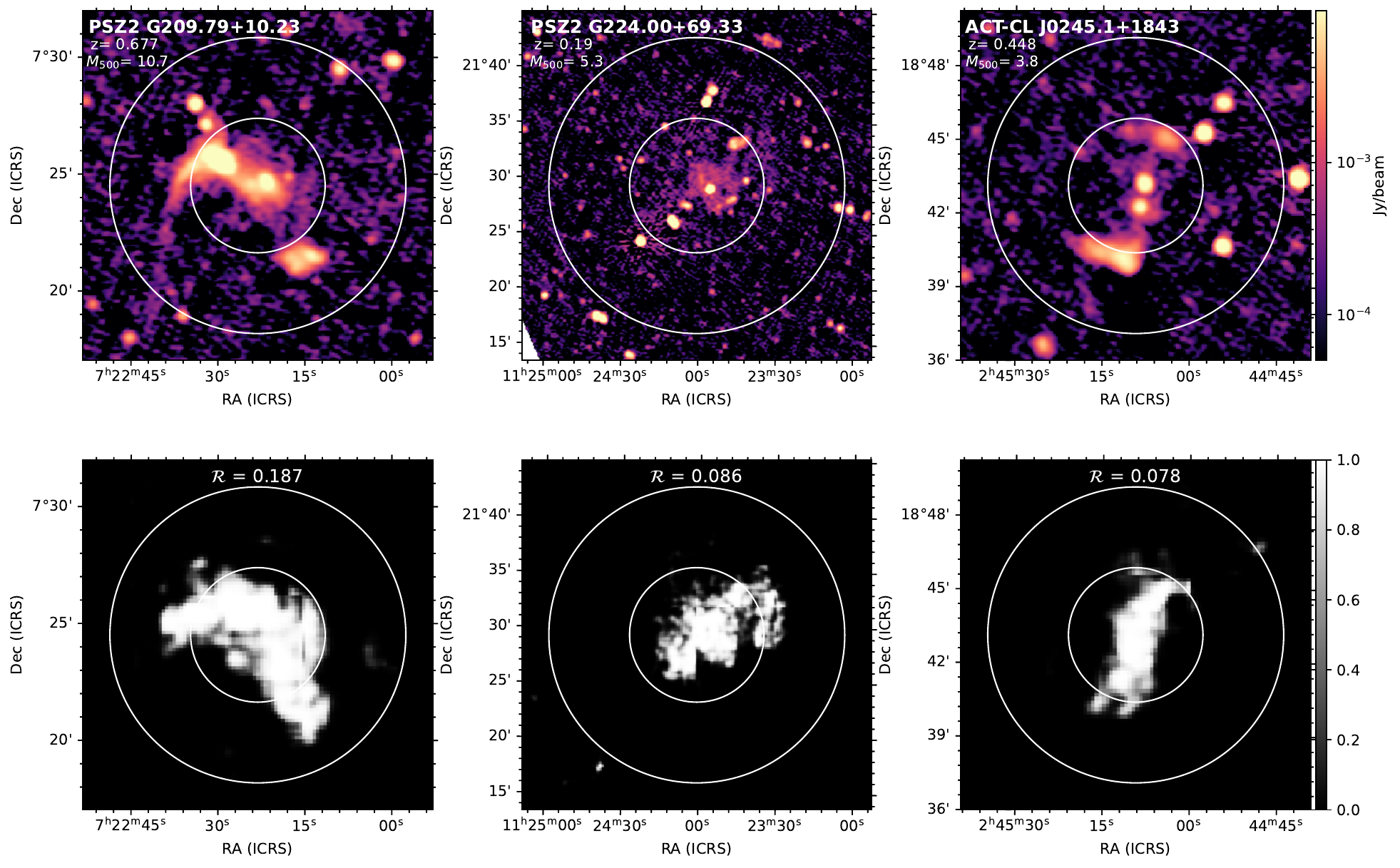}
        \caption{Example of six galaxy clusters in the curated sub-sample showing diffuse radio emission which, due to its morphology and position with respect to the cluster, could be attributed to the presence of radio halos or radio relics. The LoTSS-DR3 image at $20\arcsec$ resolution is shown in the first and third rows. The inner white circle represents the $R_{500}$ of the cluster, while the outer circle represents 2.2$R_{500}$. Redshift and $M_{500}$ (in units of $10^{14} M_{\odot}$) are shown in the image. The $\matchcal{R}$ value of each cluster is shown over the segmented maps shown in the second and fourth rows.}
        \label{fig:example}
\end{figure*}

Fig.~\ref{fig:peculiar} shows an example of a peculiar system in the curated sub-sample where diffuse radio emission was detected by the network in a region connecting three nearby galaxy clusters. In Fig.~\ref{fig:example}, we show instead some examples of galaxy clusters in the curated sub-sample without previous studies in the literature and showing diffuse radio emission, which, due to its morphology and position with respect to the cluster, could be attributed to the presence of radio halos or radio relics. 

\end{appendix}

\end{document}